\RequirePackage{fix-cm}
\documentclass[twocolumn]{svjour3}          
\usepackage{graphicx}
\newcommand{\orcid}[1]{\href{https://orcid.org/#1}{\includegraphics[width=10pt]{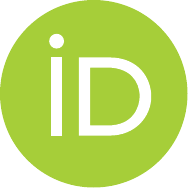}}}
\usepackage{hyperref}

\makeatletter
\expandafter\let\csname opt@amsmath.sty\endcsname\relax
\AtBeginDocument{
  \mathindent=15pt 
  \@mathmargin\@centering} 
\makeatother

\usepackage{amsmath}
\usepackage{siunitx}

\usepackage{etoolbox}
\apptocmd{\sloppy}{\hbadness 10000\relax}{}{}
\journalname{Computing and Software for Big Science}
\begin{document} 

\title{Improving Robustness of Jet Tagging Algorithms with Adversarial
Training}
\subtitle{}

\author{Annika Stein\textsuperscript{1}~\orcid{0000-0003-0713-811X} \and
        Xavier Coubez\textsuperscript{1,2}~\orcid{0000-0002-3791-2009} \and
        Spandan Mondal\textsuperscript{1}~\orcid{0000-0003-0153-7590} \and
        Andrzej Novak\textsuperscript{1}~\orcid{0000-0002-0389-5896} \and
        Alexander Schmidt\textsuperscript{1}~\orcid{0000-0003-2711-8984}
}

\institute{Correspondence to: annika.stein@cern.ch\at \\      
\textsuperscript{1} RWTH Aachen University, Aachen, Germany\\
\textsuperscript{2} Brown University, Providence, USA
}
\date{Received: 25 March 2022 / Accepted: 16 August 2022 / Published: 10 September 2022}

\maketitle
\begin{abstract}
Deep learning is a standard tool in the field of high-energy physics, facilitating considerable sensitivity enhancements for numerous analysis strategies. In particular, in identification of physics objects, such as jet flavor tagging, complex neural network architectures play a major role. However, these methods are reliant on accurate simulations. Mismodeling can lead to non-negligible differences in performance in data that need to be measured and calibrated against. We investigate the classifier response to input data with injected mismodelings and probe the vulnerability of flavor tagging algorithms via application of adversarial attacks. Subsequently, we present an adversarial training strategy that mitigates the impact of such simulated attacks and improves the classifier robustness. We examine the relationship between performance and vulnerability and show that this method constitutes a promising approach to reduce the vulnerability to poor modeling.
\keywords{High-energy physics \and Deep learning \and Jet flavor tagging \and Adversarial attacks \and Adversarial Training \and Robustness}
\end{abstract}

\section{Introduction}\label{sec:intro}
The experiments at the Large Hadron Collider (LHC) at CERN  handle large, high-dimensional datasets to find complex patterns or to identify rare signals in back\-ground-dominated regions -- tasks where machine learning and especially deep learning~\cite{cnn,Goodfellow-et-al-2016} provide considerable performance gains over traditional methods. It is expected that the relevance of new deep learning technologies will increase, with the era of the High-Luminosity LHC (HL-LHC) approaching~\cite{MLHEPwhitepaper}. However, studies with the aim of understanding a neural network's decisions demonstrate the relevance of explainability~\cite{8466590} and raise questions on the safety of systems that use artificial intelligence (AI), which is often perceived as a black-box~\cite{8466590,amodei2016concrete}. Moreover, other studies show that small modifications of the inputs (\textit{adversarial examples}) can severely affect the performance of neural networks~\cite{szegedy2014intriguing,goodfellow2015explaining} (\textit{adversarial attack}), a worrying prospect for a field that is reliant on simulation, which might be at times inaccurate. Careful exploration of the susceptibility to mismodelings is necessary to examine how severe these ``intriguing properties of neural networks''~\cite{szegedy2014intriguing} are in practice. Such effects could be driven by the fact that various popular classes of deep neural networks react linearly when exposed to linear perturbations, together with the large number of input variables~\cite{szegedy2014intriguing,goodfellow2015explaining}. As such, this property is not in conflict with a neural network's ability to approximate any function via a combination of non-linear activation functions~\cite{HORNIK1989359}, but the presence of (piecewise-)linear activation functions is sufficient to cause severe impact on performance when evaluated on first-order adversarial examples~\cite{szegedy2014intriguing,goodfellow2015explaining}. Applied to computer vision / image recognition, it has been demonstrated that modifications that involve only one pixel are enough to ``fool'' a neural network~\cite{Su_2019}.

We apply methods from AI safety~\cite{amodei2016concrete,nachman2019ai,advjets-mlhep2020} to the classification of jets based on the flavor of their initiating particle (a quark or gluon), so called jet heavy-flavor identification (tagging)~\cite{CMS-BTV-16-002,Guest_2016}. Identifying the jet flavor plays an important role in various analysis branches exploited by experiments like CMS~\cite{CMS-BTV-16-002,CMS} and ATLAS~\cite{ATLAS,ATLASbtagging}, for example, for the observation of the decay of the Higgs boson to bottom (b) quark-antiquark pairs ($\mathrm{H}\rightarrow\mathrm{b}\bar{\mathrm{b}}$)~\cite{CMS-HIG-18-016,ATLAShbb,Kogler:2018hem}. Moreover, for analyses that also apply charm (c) tagging~\cite{CMS-BTV-16-002,CMS-PAS-BTV-16-001,CMS-BTV-20-001,ATLASctagging}, such as searches for the Higgs boson decaying to c quarks~\cite{CMS-HIG-18-031,ATLAShcc,CMS-HIG-21-008}, multiclassifiers become increasingly important. Therefore, investigating the susceptibility to mismodeling could be even more relevant for c tagging. We probe the trade-off between performance and robustness to systematic distortions by benchmarking an established algorithm for jet flavor tagging with a realistic dataset. Early taggers included only the displacement of tracks as a way to discriminate heavy- from light-flavored jets, possible due to the different lifetimes of the initiating hadrons. It is also possible to leverage information related to the secondary vertices, giving rise to algorithms such as the (deep) combined secondary vertex algorithm~\cite{CMS-BTV-16-002}.

Mismodelings can arise at various steps during the Monte Carlo (MC) simulation chain, starting with the hard process (matrix element calculation), followed by the subsequent steps that model the parton shower, fragmentation and hadronization, where the perturbation order is limited, and ending with the detector simulation~\cite{CMS-BTV-20-001} which introduces imperfections such as detector misalignment and calorimeter miscalibration.\\These imperfections in the modeling, particularly for variables with high discriminating power, demand the calibration of the discriminator shapes~\cite{CMS-BTV-20-001} and call for investigations of the tagger response to slightly distorted input data~\cite{nachman2019ai}. We use adversarial attacks to model systematic uncertainties induced by these subtle mismodelings that could be invisible to typical validation methods, as proposed in Ref.~\cite{nachman2019ai}. The approach followed in this study does not eliminate these mismodelings, nor does it provide a definitive \textit{a posteriori} correction, but it helps in estimating to what extent tagging efficiency and misidentification rates could be affected~\cite{nachman2019ai,Kogler:2018hem}. We assume that more adversarially robust models also generalize better when applied to a non-training domain~\cite{Goodfellow-et-al-2016,chakraborty2018adversarial} (e.g. model evaluated on data ~\cite{nachman2019ai,CMS-BTV-20-001}).
To that end, we seek to modify the training to minimize the impact of adversarial attacks, without sacrificing  performance.

Using \textit{adversarial training}~\cite{chakraborty2018adversarial,Shaham_2018,madry2019deep} to decrease the effect of simulation-specific artefacts, we show that the injection of systematically distorted samples during the training yields a successful defense strategy. In related works, adversarial training is employed through joint training of a classifier and an adversary~\cite{Louppe:2016ylz}, making use of gradient reversal layers to connect two networks or by utilizing domain adaptation~\cite{ganin2015unsupervised,CMS-EXO-19-011,Ciprijanovic:2021xsy,Babicz:2022zgm}. Other approaches towards regularization and generalization in the realm of high-energy physics include data augmentation or uncertainty-aware learning~\cite{ghosh2021uncertainty}.
\section{Dataset and Input Features}\label{sec:data}
We use the \mbox{\textsc{Jet Flavor dataset}}~\cite{Guest_2016}. These samples are generated with  \mbox{\textsc{Madgraph5}}~\cite{madgraph5} and \mbox{\textsc{Pythia 6}}~\cite{Sjostrand:2006za}. The detector response is simulated with \mbox{\textsc{Delphes 3}}~\cite{deFavereau:2013fsa}, using the ATLAS~\cite{ATLAS} detector configuration.\\Jets are clustered with the anti-$k_\text{T}$ algorithm~\cite{Cacciari_2008} using the \mbox{\textsc{FastJet}}~\cite{Cacciari:2011ma} package, with $R=0.4$. 
Secondary vertices are reconstructed using the adaptive vertex reconstruction algorithm, as implemented in RAVE~\cite{Waltenberger:2011zz}. Parton matching within a cone of $\Delta R < 0.5$ is used to define the simulated truth labeling of jets. The targets fall in one of the three classes, depending on the jet flavor: light (up, down, strange quarks or gluons), charm, or bottom~\cite{Guest_2016}, where the heavier flavor takes precedence in case multiple partons are found. Using this hierarchy for light, charm and bottom, the flavor content is distributed among the classes as $48.7\%:12.0\%:39.3\%$.
\subsection{Input Features}\label{subsec:input_features}
A description of all input variables is given in Tables~\ref{tab:highlevel_inputs} and~\ref{tab:lowlevel_inputs}, and is based on Ref.~\cite{Guest_2016}; here we only summarize the main categorization.

Input features are organized hierarchically. Low-level features consist of tracks and their helix parameters, along with the track covariance matrix. Additional information is taken from the relationship between each track and the associated vertex. Up to 33 tracks, sorted by impact parameter significance, are available per jet, however, we only consider the first six.

At jet level, expert (high-level) features are constructed as a function of the low-level inputs, for example by summing over all tracks or summing over secondary vertices, such as the weighted sum of displacement significances. Additionally, kinematic features of the jet are taken into account.

Missing or otherwise unavailable variables are filled with a convenient default value for later processing.
\subsection{Preprocessing}\label{subsec:preprocessing}
The entire dataset consists of $11,491,971$ jets, which are split randomly into training ($72\%$), validation ($8\%$) and test ($20\%$) sets. Input features are normalized such that they have a mean of 0 and standard deviation of 1. The scaling is calculated only using the training dataset distributions, excluding the defaulted values.
Defaulted input values are set just below the minima of the primary input distributions ensuring no interference between regular and irregular (or missing) values. Minimizing the gap between the default value to the rest of the distributions improves training convergence. 
This technique of missing data imputation allows us to create fixed length input shapes that are transferred to the first layer of a deep feed-forward neural network, and at the same time prevents vanishing or exploding gradients due to extreme values for the defaults~\cite{Goodfellow-et-al-2016,stevens2020deep,kuhn2019feature}.

Sample weights are calculated to exclude a potential flavor dependence of the classifier on the particular kinematic properties of the chosen dataset and to correct for the inherent class imbalance. The reweighting aims at identical, kinematic distributions for all three flavors and is done with respect to the jet transverse momentum ($p_\text{T}$) and pseudorapidity ($\eta$) distributions~\cite{CMS-BTV-16-002}. The target shape is the average of the three initial distributions, thus balancing the relative fractions for the three classes at the same time. These distributions are binned into a 2D grid of $50\times 50$ bins, spanning ranges between $(20,900)~\si{\GeV}$ and $(-2.5,2.5)$, respectively. When calculating the loss per batch, these weights are multiplied to the individual losses per sample.
\section{Methods}\label{sec:methods}
\subsection{Reference Classifier}\label{subsec:reference_classifier}
The studies are carried out on a jet flavor tagging algorithm similar in implementation to the ones used at the LHC experiments, such as ATLAS and CMS. We use a fully-connected sequential model with five hidden layers of $100$ nodes each. We use dropout layers~\cite{dropout} with a $10\%$ probability of zeroing out each neuron at each hidden layer to prevent overfitting. The Rectified Linear Unit (ReLU) activation function~\cite{Goodfellow-et-al-2016,CMS-BTV-16-002,stevens2020deep} is used for the hidden layers, the activation of the output layer is computed with the Softmax~\cite{Goodfellow-et-al-2016,CMS-BTV-16-002} function. In total, there are 184 input nodes, where the low-level per track features are flattened. We define three output classes, analogous to the dataset. 

As loss function, we use the categorical cross entropy loss~\cite{stevens2020deep,pytorchCrossEntropy}, multiplied with an additional term that downweights easy-to-classify samples during training. The resulting formula for the so called focal loss~\cite{lin2018focal,multiclassfocallossCODE,CMS:2022mqo} evaluated for one batch of length $N$ is given as: 
\begin{align}\frac{1}{\sum_{i=1}^N w_i} \sum_{i=1}^N w_i \sum_{j=1}^3 - (1-y_{ij})^\gamma \cdot \hat{y}_{ij} \log(y_{ij}),
\end{align} where $y_{ij}$ is a placeholder for the output probability assigned to one of the three possible flavors $j$ of the jet $i$, $\hat{y}_{ij}$ can be understood as the one-hot-encoded truth label which is either 0 or 1, $w_i$ is the sample weight obtained from preprocessing and $\gamma$ is called focusing parameter. Though we already treat the class imbalance by reweighting the nominal loss function, without the focusing term, the neural network is prone to assign the most frequent class. In a setting with highly-imbalanced data the chosen technique ensures smooth classifier output distributions, which we achieve by choosing a focusing parameter of $\gamma=25$.

Model parameters are updated with the Adaptive Moments Estimation (Adam) optimizer~\cite{adam} using PyTorch's~\cite{paszke2019pytorch} default settings, which is further controlled with a learning rate schedule~\cite{253713} that starts at $0.0001$ and decays proportionally to $\left(1+\frac{\text{epoch}}{30}\right)^{-1}$. The batch size has been fixed to $2^{16}=65{,}536$. To ensure that there is no overfitting, training is stopped when the validation loss no longer improves~\cite{Goodfellow-et-al-2016}. For each training, the model's parameters are saved after each iteration through the full training dataset (i.e. after each epoch) to store a checkpoint for later evaluation.
\subsection{Evaluation Metrics}\label{subsec:evaluation_metrics}
While multi-class taggers are convenient for implementation, for physics analysis purposes, one is often interested in constructing classifiers distinguishing two classes at a time. We take appropriate likelihood ratios of the bottom, charm and light output classes as needed for discrimination. The likelihood ratio XvsY for discriminating class X from Y is given as:
\begin{gather}
\frac{P(\text{X})}{P(\text{X}) + P(\text{Y})}.
\end{gather}
For example, for the BvsL discriminator, $P(\text{X})$ and $P(\text{Y})$ refer to the classifier's score for the bottom and light flavor jets, respectively. The performance of the binary classifiers is visualized and evaluated using Receiver Operating Characteristic (ROC) curves~\cite{10.1145/1143844.1143874,powers2008,5978225}. With some loss of information, a ROC curve is characterized by its area under the curve (AUC), which can be used as a reasonable single scalar proxy for the classifier performance~\cite{branco2015survey}. It should be noted that due to a large class imbalance in the available dataset, accuracy could be an inaccurate measure of the performance~\cite{branco2015survey}.
\subsection{Adversarial Attacks}\label{subsec:adversarial_attacks}
One way to generate adversarial inputs is the Fast Gradient Sign Method (FGSM)~\cite{Goodfellow-et-al-2016,goodfellow2015explaining}, which modifies the inputs in a systematic way, such that the loss function increases. First, the direction of the steepest increase of the loss function around the raw inputs is computed. Mathematically, the operator that allows to retrieve the ``steepest increase'' is the gradient of the loss function with respect to the inputs. Once the direction is known, of which only the sign is kept, this vector is multiplied with a (small) limiting parameter $\epsilon$ to specify the desired severity of the impact. Then, the nominal inputs are shifted by this quantity. It can, therefore, be seen as a technique to maximally disturb the inputs or maximally confuse the network without necessarily manifesting in the input variable distributions.

Expressed in a single equation, the FGSM attack generates adversarial inputs $x_\text{FGSM}$ from raw inputs $x_\text{raw}$ by computing
\begin{align}\label{eq:fgsm}
	x_\text{FGSM} = x_\text{raw} + \epsilon \cdot \mathrm{sgn}\left(\nabla_{x_\text{raw}}J(x_\text{raw},y)\right),
\end{align}
where $\mathrm{sgn}(\alpha)$ stands for the sign of $\alpha$. In Eq.~(\ref{eq:fgsm}), the loss function is denoted as $J(x_\text{raw},y)$, a function of the inputs ($x_\text{raw}$) and targets ($y$). Moreover, the FGSM attack can be interpreted as a method that locally inverts the approach of gradient descent by performing a gradient ascent with the loss function, but in the input space~\cite{goodfellow2015explaining,chakraborty2018adversarial,madry2019deep}. Using the terminology of Ref.~\cite{chakraborty2018adversarial}, this is a white box attack with full knowledge of the network (architecture and parameters).

\begin{figure}[ht]%
	\centering
	\includegraphics[width=0.45\textwidth]{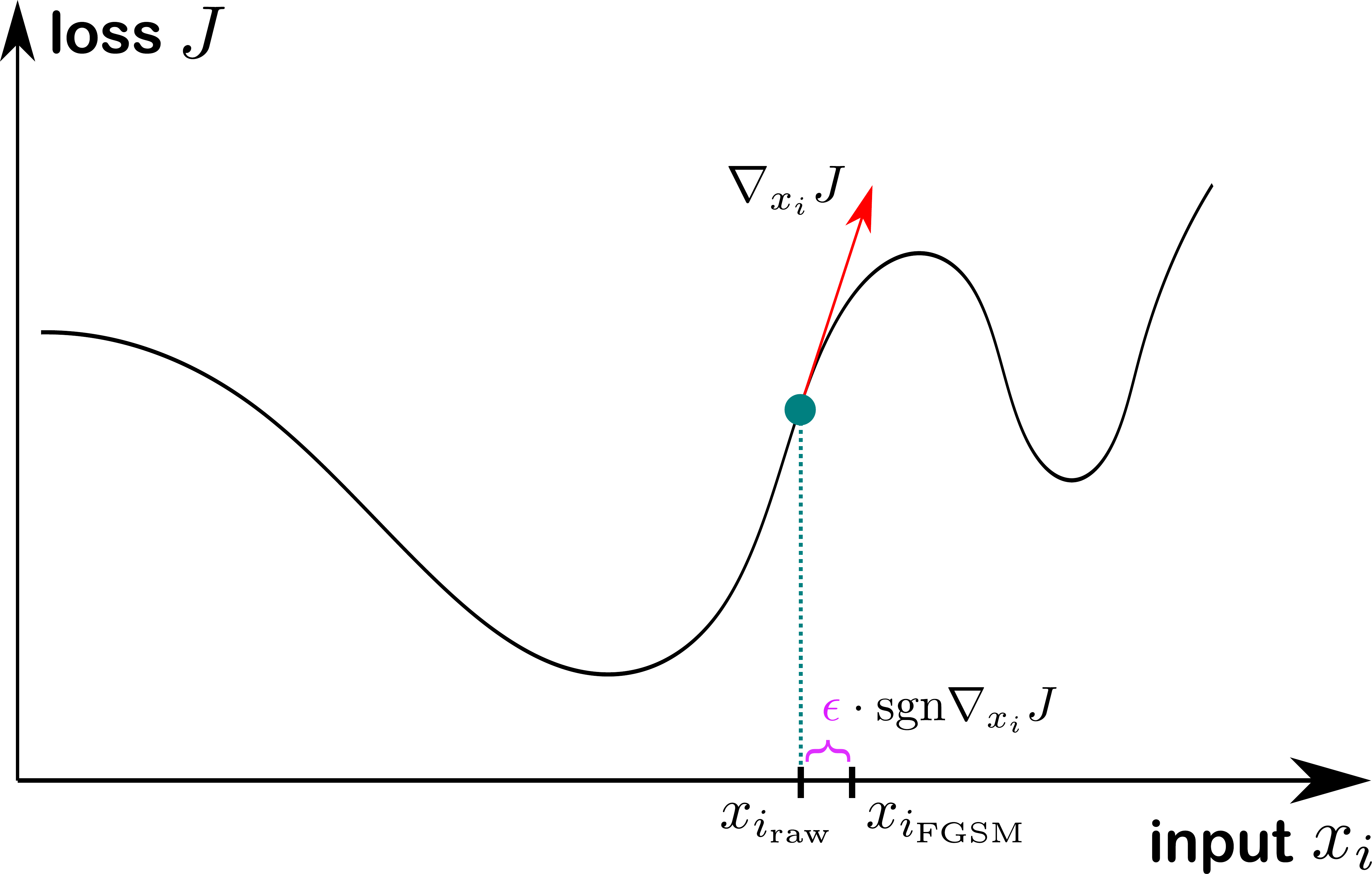} %
	\caption[Visualization of the generation of adversarial inputs by applying the FGSM attack.]{Visualization of the generation of adversarial inputs by applying the FGSM attack.}%
	\label{graphic:fgsm}%
\end{figure}

The corresponding visualization is shown in Fig.~\ref{graphic:fgsm}, however, for didactic reasons with one input variable $x_i$ only. In practice, this method is applied multidimensionally, assigning the same limiting parameter $\epsilon$ in each input dimension.
Whereas the gradient of an arbitrary function could yield any value, the distortion should stay in reasonable bounds to mimic the behaviour of possible mismodelings or differences between data and simulation~\cite{goodfellow2015explaining,nachman2019ai}. Therefore, we go only a small step in the direction of the gradient, which is expected to introduce practically unnoticeable changes of the input distributions~\cite{szegedy2014intriguing,goodfellow2015explaining}.

Increasing the number of inputs to the model also increases the susceptibility towards adversarial attacks, because each shift by $\epsilon$ for additional features is propagated to the change in activation~\cite{goodfellow2015explaining}. Thus it is conceivable that individual feature distributions remain almost unaffected, but the performance of the neural network is substantially deteriorated.

The FGSM attack does not necessarily replicate a global worst-case scenario~\cite{madry2019deep}. Depending on the actual properties of the loss surface, the adversarial attack could shift the inputs also into local minima (or at least harmless regions), if the limiting parameter is chosen unluckily. On average, with small distortions only, it is still expected that in a given region, the attack will maximally confuse the model up to first order.

In this implementation, the FGSM attack is not applied to integer variables, such as the number of tracks, and defaulted values, which would not be shifted by $\epsilon$ in a physically meaningful way. 

As large distortions of input variables would be easy to detect, a limit of $25\%$ with respect to the original value is applied on the perturbation. The modified value $x_\text{FGSM}$ is then given by Eq.~(\ref{eq:fgsm_mod}), where $x$ denotes the original input value, $x'$ the transformed (preprocessed) value and $\epsilon$ the FGSM scaling factor. Inverting the normalization is denoted by $()^{-1}$.
\begin{align}
	\nonumber x_\text{FGSM} = \Big( \Big. x~+ &~\mathrm{sgn}\left(\nabla_{x}J(x,y)\right) 
	\cdot \min\big\{\big.\\ \nonumber
	&|(x' + \mathrm{sgn}\left(\nabla_{x}J(x,y)\right) \cdot \epsilon)^{-1} - x|,\\ &|0.25 \cdot x|\big.\big\} \Big. \Big) \,'
	\label{eq:fgsm_mod}
\end{align}
Distortions of low-level features are not propagated to high-level features, instead each feature is taken into account via the multidimensional gradient only. Therefore, correlations are not fully taken into account.
\subsection{Adversarial Training}\label{subsec:adversarial_training}
\begin{figure*}[ht]
	\centering
	\includegraphics[width=0.9\textwidth]{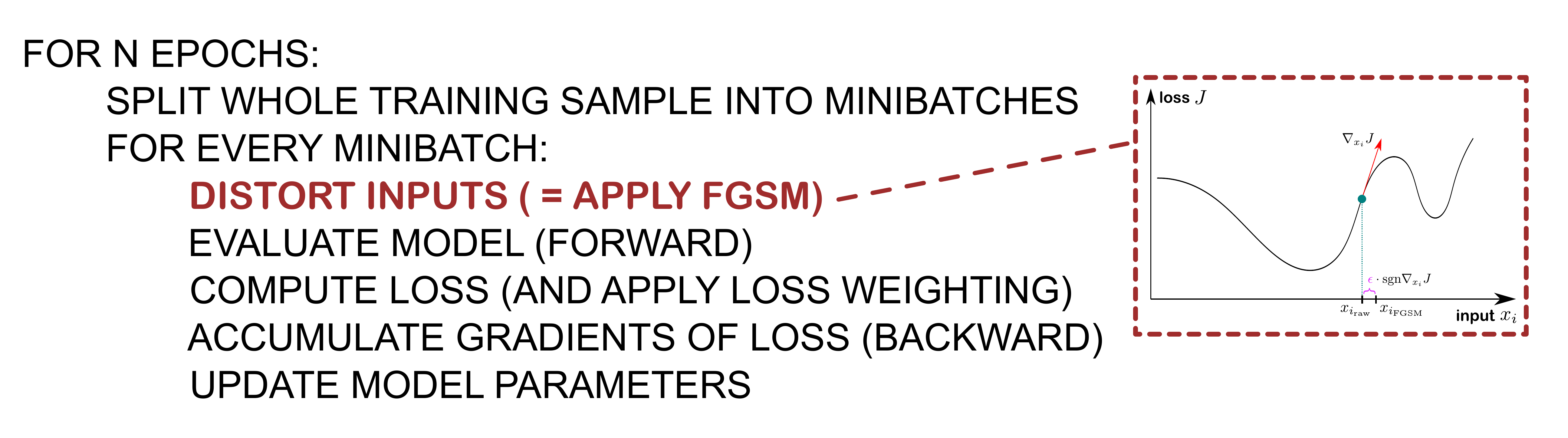}
	\caption[Adversarial training algorithm.]{Adversarial training algorithm. The inputs are distorted prior to the forward and backward passes, with the FGSM attack. The standard training algorithm denoted in black is based on Ref.~\cite{stevens2020deep}, the modified implementation for adversarial training is demonstrated in Ref.~\cite{advjets-mlhep2020}.}%
	\label{graphic:adversarial_training}%
\end{figure*}

The approach that will be followed in this study is a simple type of adversarial training that injects perturbed inputs already during the training phase~\cite{chakraborty2018adversarial}. The algorithmic description is shown in Fig.~\ref{graphic:adversarial_training}. The difference to the nominal and adversarial training is highlighted in red.
\begin{figure*}[ht]
	\centering
	\includegraphics[width=0.9\textwidth]{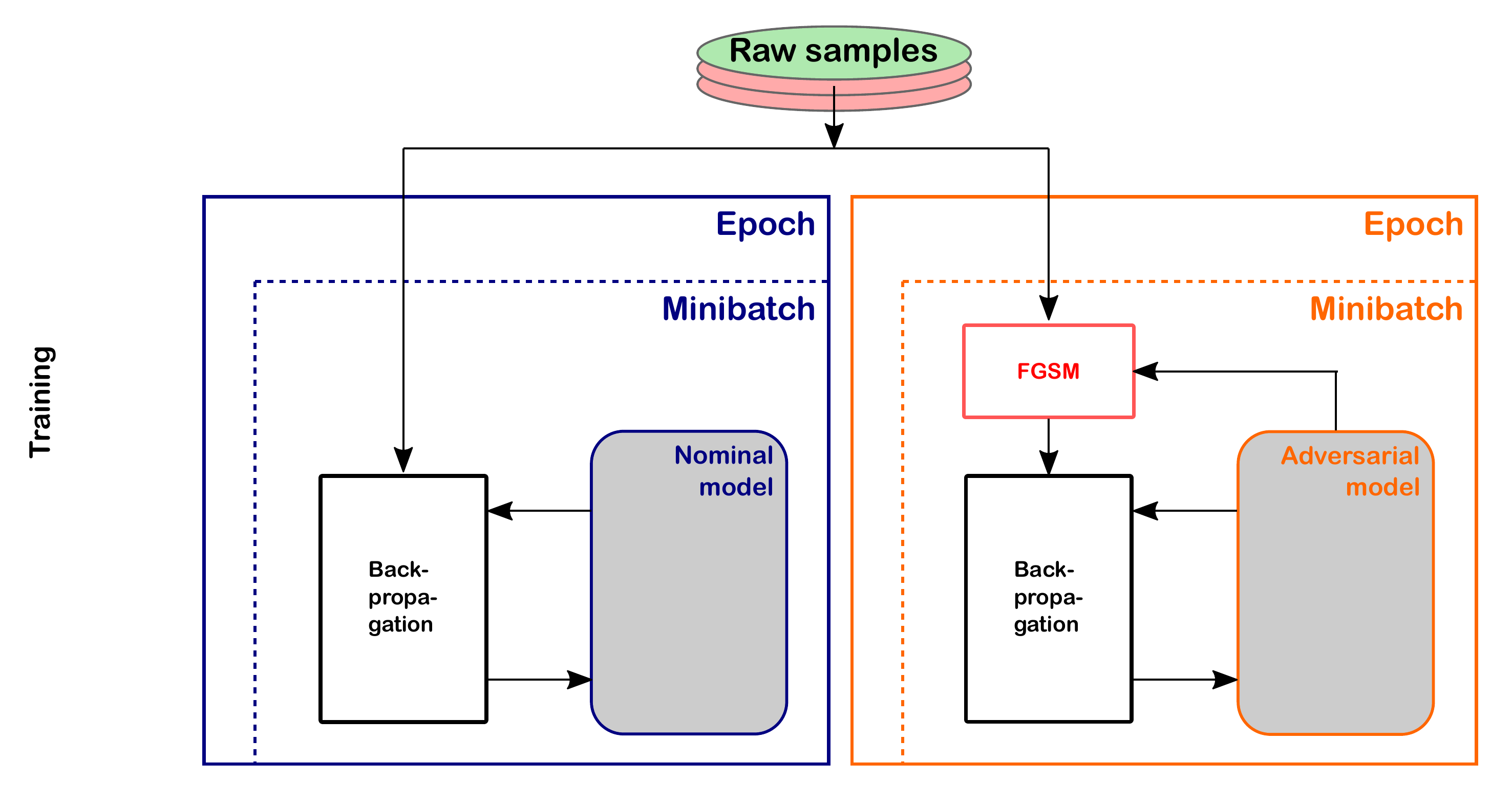}
	\caption[Comparison of the nominal and adversarial training against the FGSM attack.]{Comparison of the nominal and adversarial training against the FGSM attack.}%
	\label{graphic:training_nominal_adversarial}%
\end{figure*}
In fact, in this approach the neural network never sees the raw inputs during the whole training step~\cite{chakraborty2018adversarial,Shaham_2018,madry2019deep}. In Fig.~\ref{graphic:training_nominal_adversarial}, this is shown with the insertion of a red block prior to backpropagation. The idea is that by applying the FGSM attack continuously to the training data (for every minibatch, i.e. with every intermediate state of the model after updating the model parameters), the network is less likely to learn the simulation-specific properties of the used sample. Instead, the introduction of a saddle point into the loss surface is expected to improve the generalization capability of the network~\cite{Goodfellow-et-al-2016,chakraborty2018adversarial,madry2019deep}. This can be understood as a ``competition'' between gradient descent to solve the outer minimization problem and gradient ascent to handle the inner maximization~\cite{madry2019deep}.

Madry et al.~\cite{madry2019deep} have shown that this is an effective method to reduce susceptibility to first-order adversaries, obtained from an FGSM attack. In that sense, adversarial training could also be described as a regularization technique, but a more systematic one than only randomly smearing inputs (another example of data augmentation), randomly deleting connections (dropout), or assigning a probability to the different targets to be wrong (label smoothing)~\cite{Goodfellow-et-al-2016}.
\begin{figure*}[ht]
	\centering
	\includegraphics[width=0.9\textwidth]{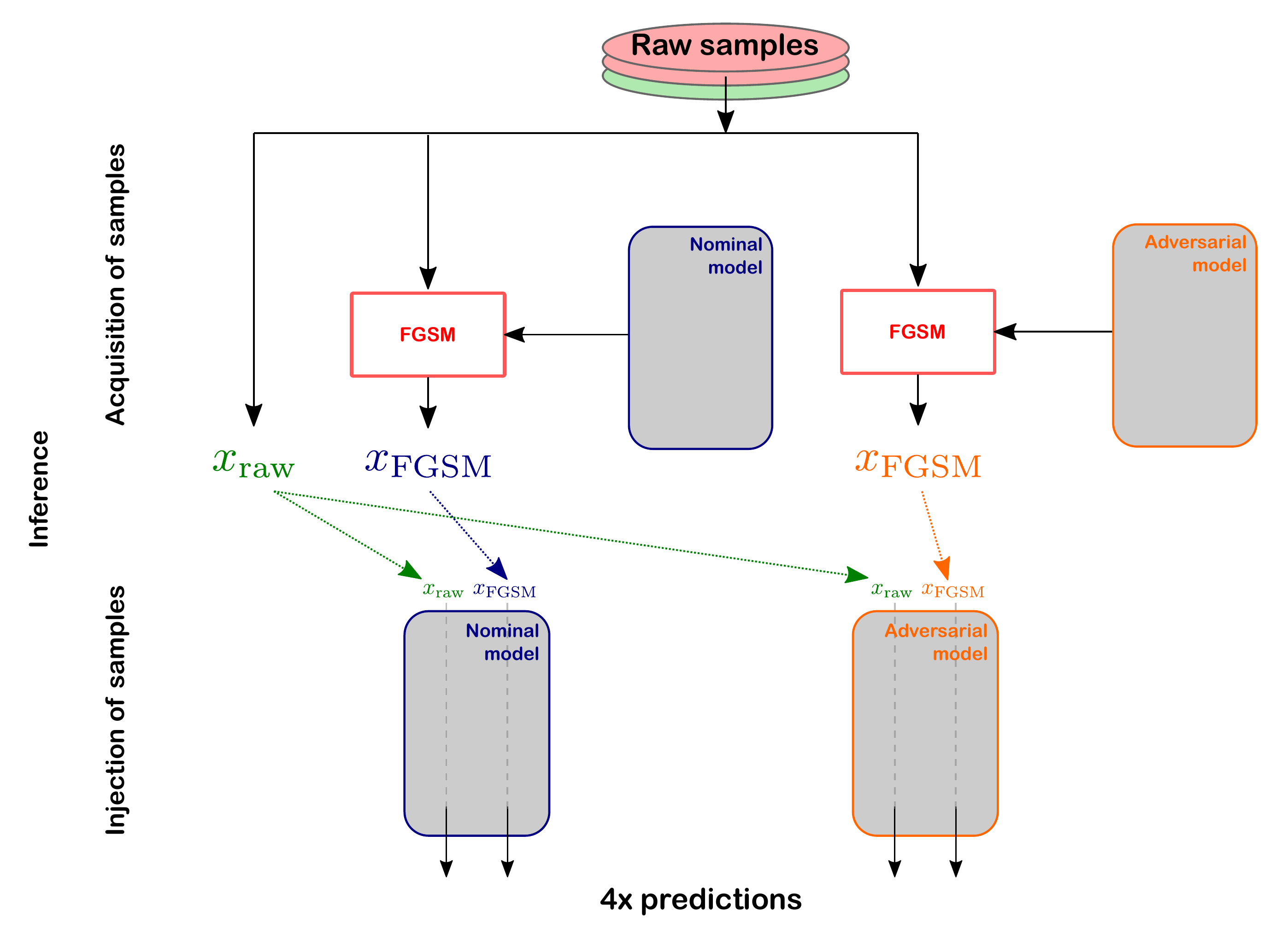}
	\caption[Schematic overview of the inference process when performing a comparison of robustness of both training strategies.]{Schematic overview of the inference process when performing a comparison of robustness of both training strategies. Evaluation of the nominal training (green and blue paths) is described in Sect.~\ref{subsec:experiments_nominal_training}, while the comparison for the adversarial training, including all four combinations is described in Sect.~\ref{subsec:experiments_adversarial_training}.}%
	\label{graphic:inference_nominal_adversarial}%
\end{figure*}

The principle behind this technique involves the linearity of neural networks to which the high susceptibility to mismodelings is attributed. Adversarial training can be interpreted as a method that adjusts the loss surface to be locally constant around the inputs and that downsizes the impact of perturbations evaluated with a high-dimensional linear function~\cite{Goodfellow-et-al-2016}. Slightly distorted inputs then cannot significantly increase the value of the loss function, because it is almost flat in the vicinity of the raw inputs~\cite{fawzi2016robustness}. This can be seen as a geometrical problem where the loss manifold is flattened~\cite{fawzi2016robustness,fawzi2016analysis,li2018visualizing,fort2020deep}. When evaluating this adversarially-trained model with distorted test inputs, the model should be more robust to those modifications and the performance should not be affected as much as with the generic training. The price for the increased robustness is that the maximally achievable performance on raw inputs can be somewhat reduced with respect to the nominal training~\cite{Goodfellow-et-al-2016}. During adversarial training, the FGSM attack uses $\epsilon=0.01$ when injecting adversarial samples, and no further restrictions are applied, i.e. there is no limitation of the attack with respect to the relative scale of the impact on different values and Eq.~(\ref{eq:fgsm}) holds.
\subsection{Inference}\label{subsec:inference}
The inference step is split into two separate parts, which can be seen in Fig.~\ref{graphic:inference_nominal_adversarial}. First, the relevant samples need to be acquired. These can be either original (raw) samples or systematically distorted samples. Both trainings under consideration have their own respective loss surfaces, which continuously change during the training process. Therefore, samples that maximally deteriorate the performance of one model do not necessarily confuse another model. To cause a severe impact, the FGSM attack will be applied individually per training. A similar argument can be made for different checkpoints of the training, where we also craft adversarial samples per epoch to reflect the model's exact status and loss surface. After a fixed number of epochs or after convergence of both training strategies, this yields three different sets of samples: nominal samples (green, equal for both contenders), FGSM samples corresponding to the nominal training (blue), and FGSM samples that have been created for the adversarial training (orange). These can then be injected into the different models for evaluation.
\begin{figure*}[ht]
	\makebox[\textwidth][c]{
	\centering
	\includegraphics[height=0.3\textwidth]{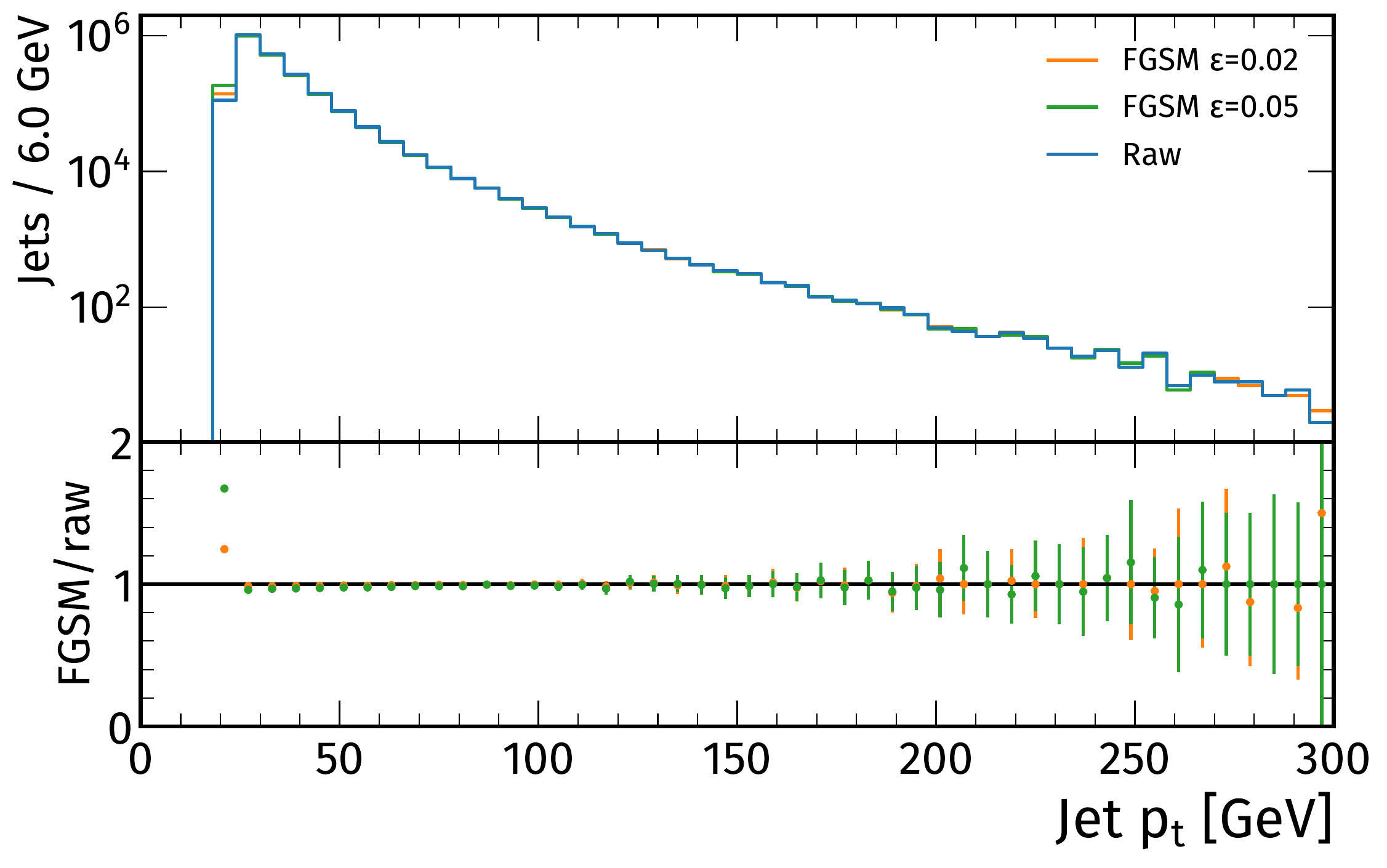} %
	\includegraphics[height=0.3\textwidth]{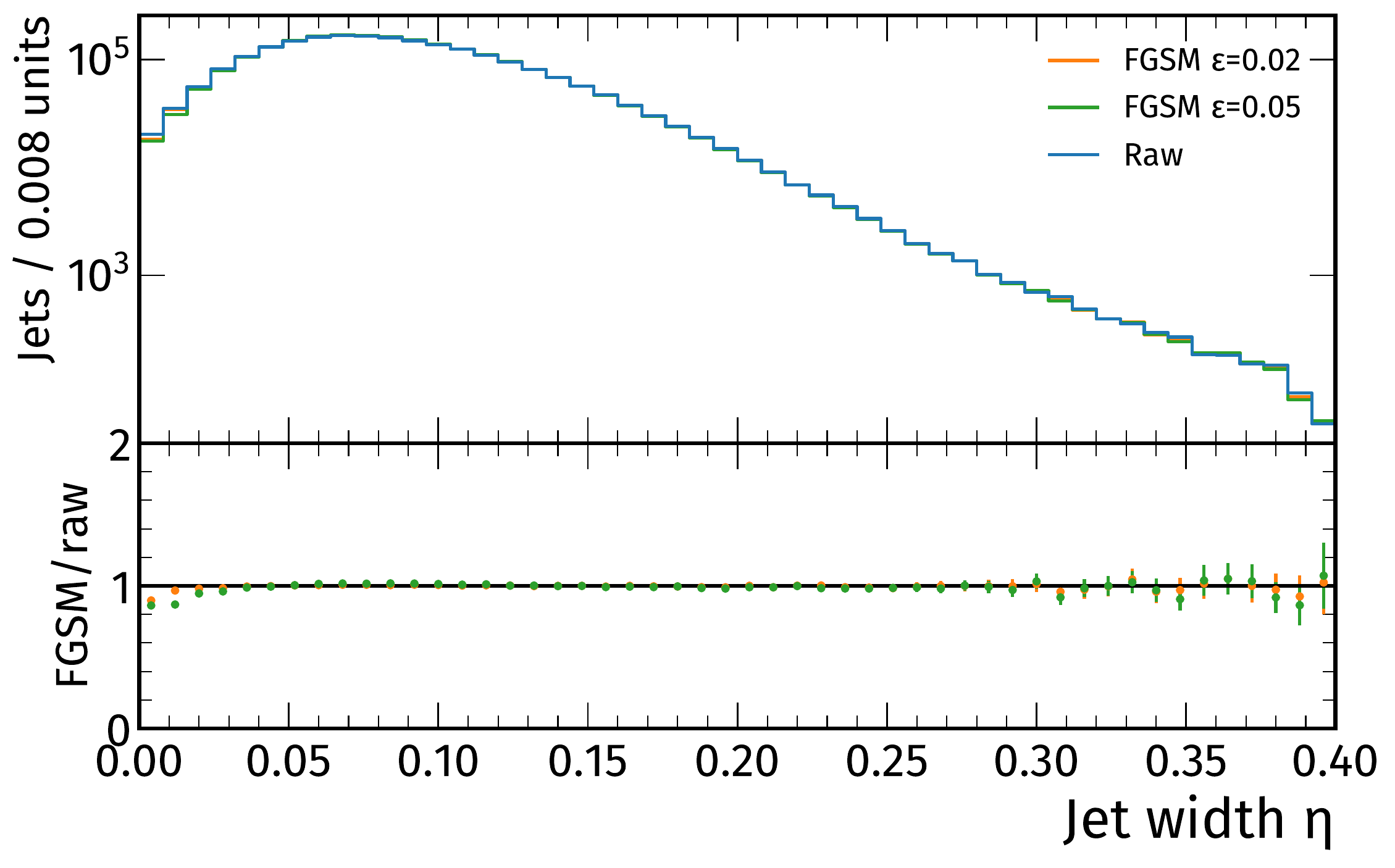}
} \\ %
\makebox[\textwidth][c]{
	\includegraphics[height=0.3\textwidth]{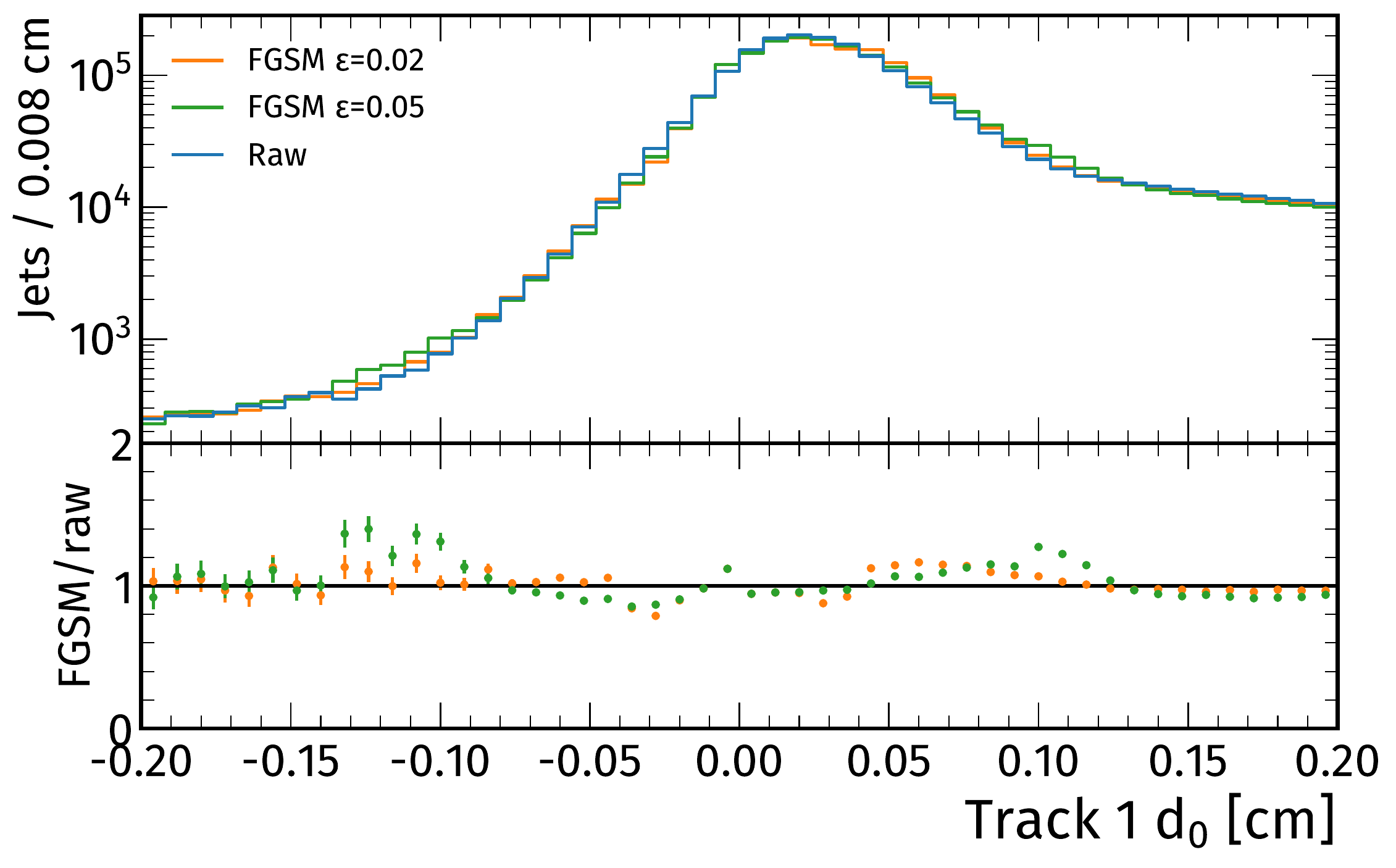} %
	\includegraphics[height=0.3\textwidth]{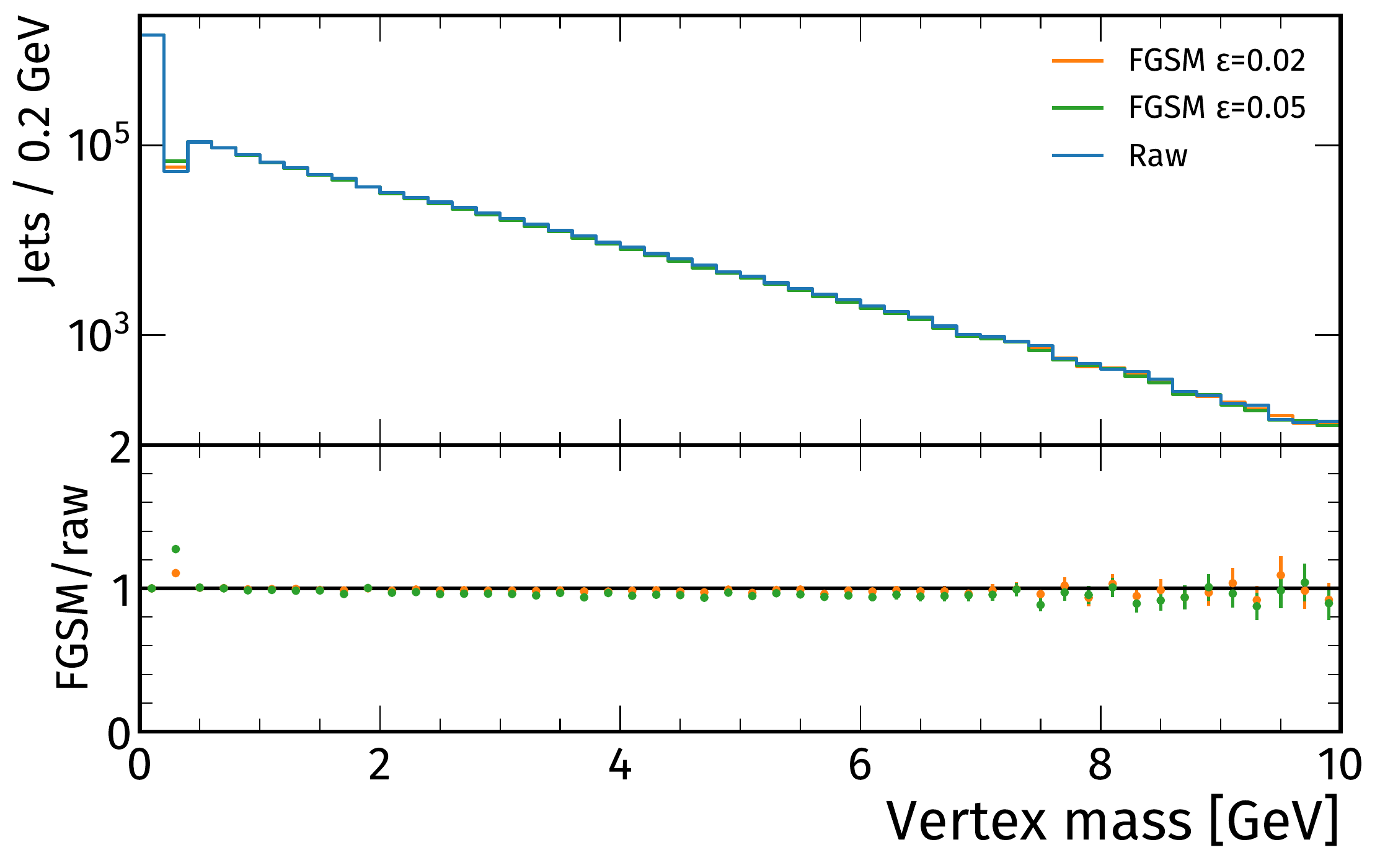}
} %
	\caption[Distributions of raw and systematically distorted inputs, for a set of features containing high- and low-level information.]{Distributions of raw and systematically distorted inputs, for a set of features containing high- and low-level information. The displayed range for the signed impact parameter ($d_0$) of the first track has been clipped to the most relevant central region, where distortions naturally appear enhanced.}%
	\label{plot:inputs_summed_flav}%
\end{figure*}
\section{Robustness to Mismodeling}\label{sec:experiments_robustness}
\subsection{Adversarial Attack}\label{subsec:experiments_adversarial_attack}
As we are interested in producing disturbances that would simulate the behaviour of systematic uncertainties, we verify that the distorted distributions remain within an envelope expected by the typical data-to-simulation agreement. The effect of the FGSM attack at two values of $\epsilon$ compared to the nominal distribution is shown for four input variables (both high and low-level inputs) in Fig.~\ref{plot:inputs_summed_flav}. Even with the largest value of $\epsilon = 0.05$ chosen for the following performance studies, the modifications of input shapes remain marginal, within typical data-to-simulation agreements of the level of 10--20\%~\cite{CMS-BTV-16-002}.
\subsection{Vulnerability of the Nominal Training}
\label{subsec:experiments_nominal_training}
\begin{figure*}[ht]%
	\makebox[\textwidth][c]{
	\centering
	\includegraphics[width=0.45\textwidth]{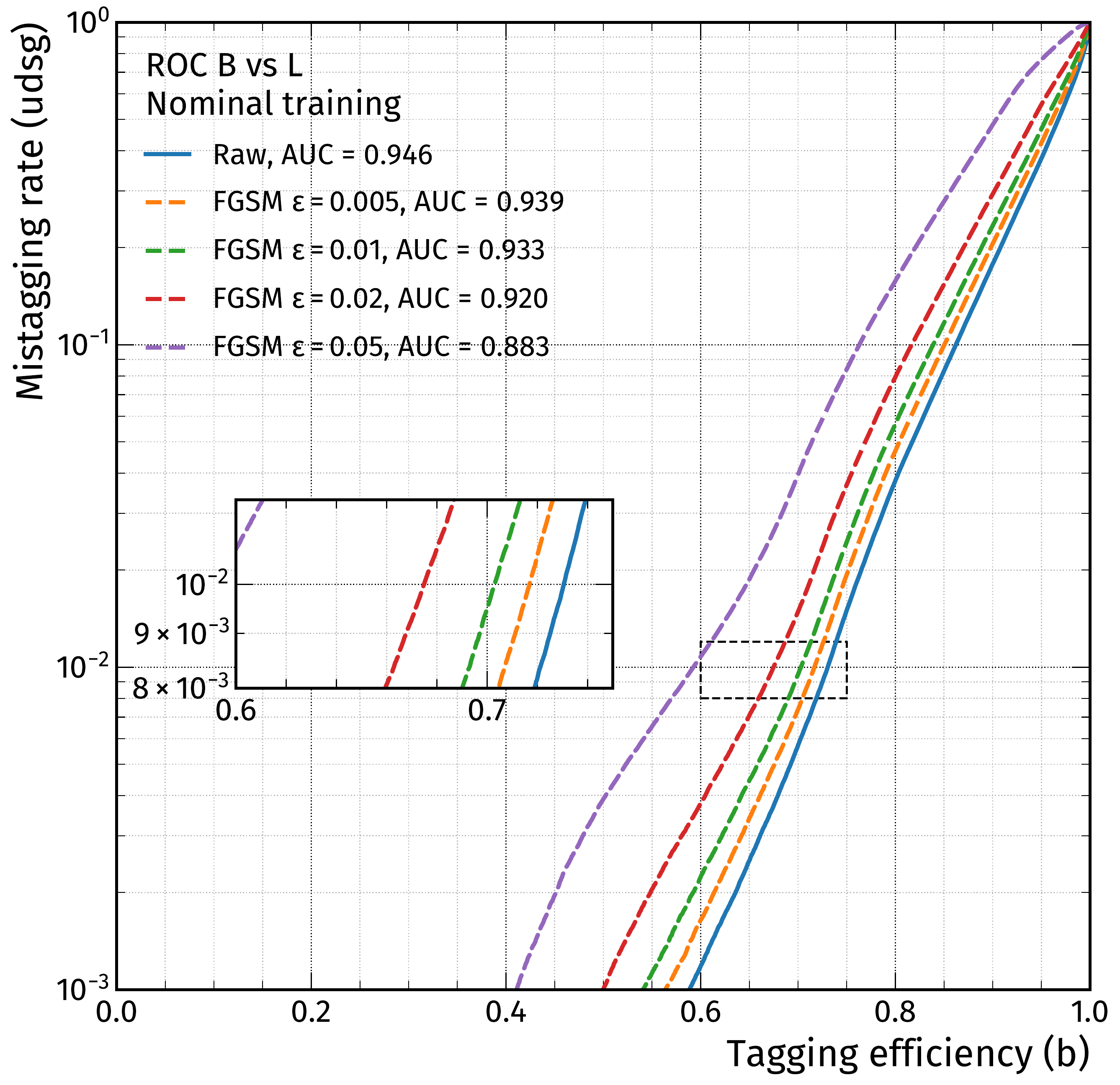}
	\includegraphics[width=0.45\textwidth]{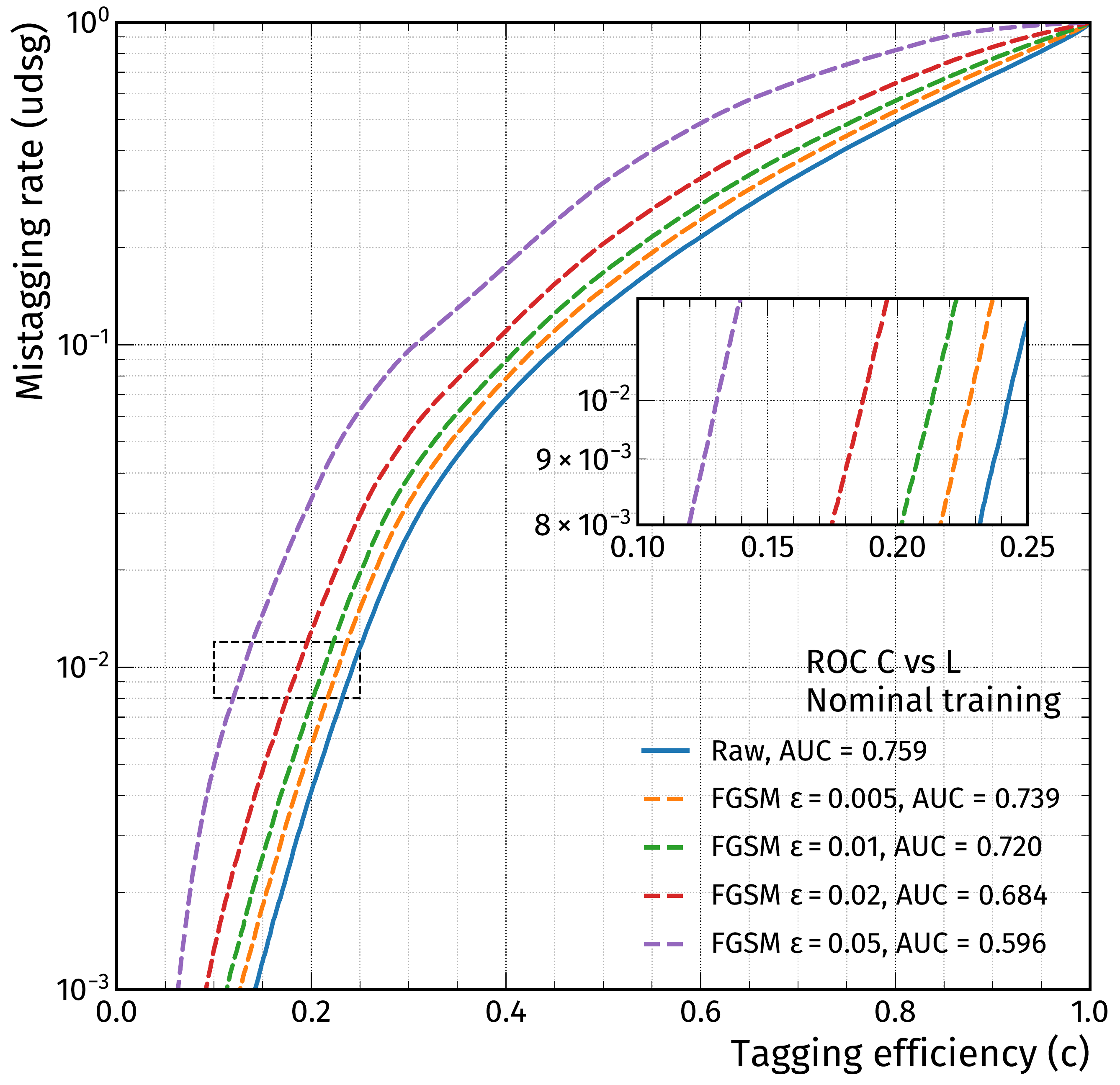}
} %
	\caption[ROC curves for the BvsL (left) and CvsL discriminator (right), using the nominal training and applying FGSM attacks of different magnitudes.]{ROC curves for the BvsL (left) and CvsL discriminator (right), using the nominal training and applying FGSM attacks of different magnitudes. The model is evaluated when the training has reached peak performance.}%
	\label{plot:basic_ROC}%
\end{figure*}

First, we establish how susceptible the nominal model is to the FGSM attack (mismodeling) of various magnitudes. Figure~\ref{plot:basic_ROC} shows the ROC curves for the BvsL (left) and CvsL (right) discriminators, on FGSM data\-sets generated with varying parameter $\epsilon$ and on the nominal inputs. As expected, the model performs best on undisturbed test samples with AUC of $0.946$, but the performance decays quite quickly with increasing $\epsilon$. At $\epsilon=0.05$, which still only causes barely visible differences in the input distributions, the model reaches AUC of $0.883$. At 1\% mistag working point, this would correspond to a decrease in signal efficiency from 73 to 60\%, requiring a scale factor of 0.82.

\begin{figure*}[ht]%
	\makebox[\textwidth][c]{
	\centering
	\includegraphics[width=0.45\textwidth]{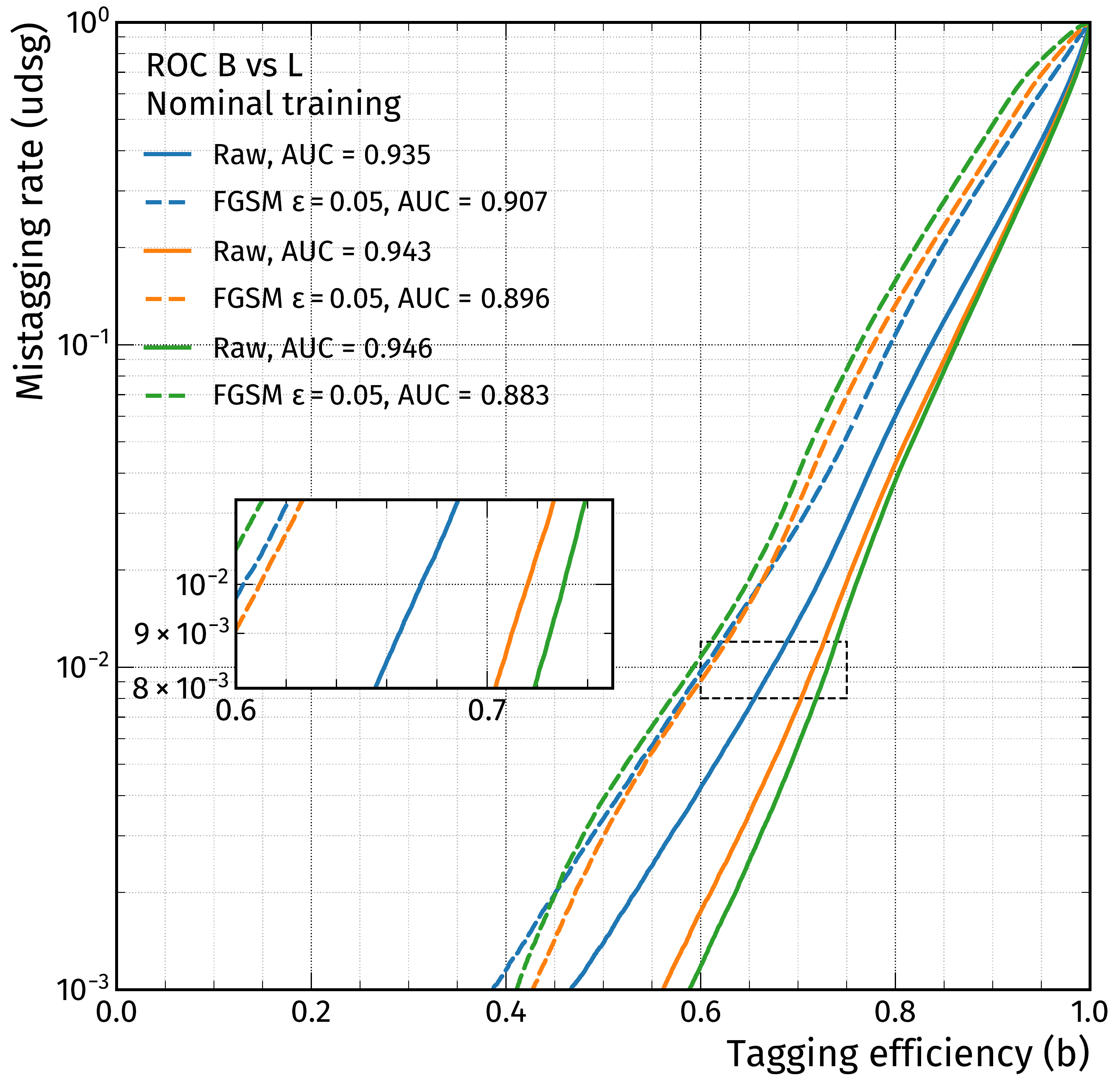}
	\includegraphics[width=0.45\textwidth]{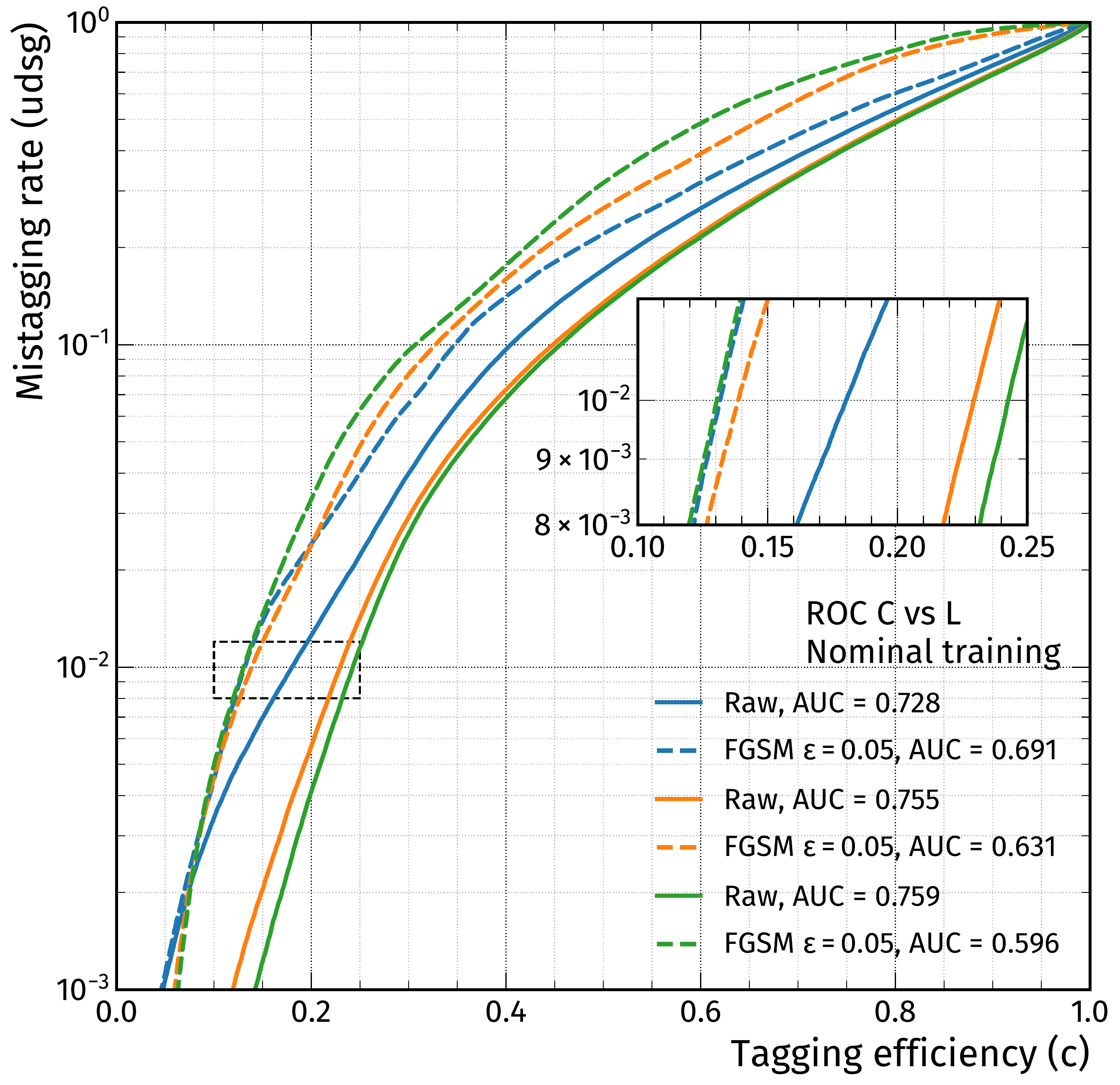}
} %
	\caption[ROC curves for the BvsL (left) and CvsL discriminator (right), using the nominal training and applying FGSM attacks with $\epsilon=0.05$ at various checkpoints of the training that each come with different nominal performance.]{ROC curves for the BvsL (left) and CvsL discriminator (right), using the nominal training and applying FGSM attacks with $\epsilon=0.05$ at various checkpoints of the training that each come with different nominal performance. Solid lines in different colors represent nominal performance gain with an increased number of epochs, dashed lines show corresponding performance on individually crafted FGSM samples for the particular checkpoints.}%
	\label{plot:basic_ROC_checkpoints}%
\end{figure*}

In the context of the ongoing hunt for better performing classifiers, it is of interest to investigate the susceptibility in relation to the performance. Some insight can be gleaned by evaluating the performance of the classifier at various steps during the training on both the nominal and the perturbed datasets with a fixed $\epsilon=0.05$, where an AUC value is calculated for each checkpoint. This dependence is shown in Fig.~\ref{plot:basic_ROC_checkpoints}, again for the two discriminators. Not surprisingly, before the training performance becomes saturated, longer training leads to an increase in nominal performance. However, at the same time it shows higher vulnerability towards adversarial attacks. In fact the performance on the perturbed datasets follows exactly the opposite trend. Another way to phrase this finding is that the least performant configuration (after only few epochs or iterations through the full training dataset) shows the highest robustness, i.e. the gap between dashed and solid lines is minimal.
\subsection{Improving Robustness Through Adversarial Training}
\label{subsec:experiments_adversarial_training}
In this subsection, the studies described above are repeated with the adversarial model, using the same setup for the attacks when performing the inference.

\begin{figure*}[ht]%
	\makebox[\textwidth][c]{
	\centering
	\includegraphics[width=0.45\textwidth]{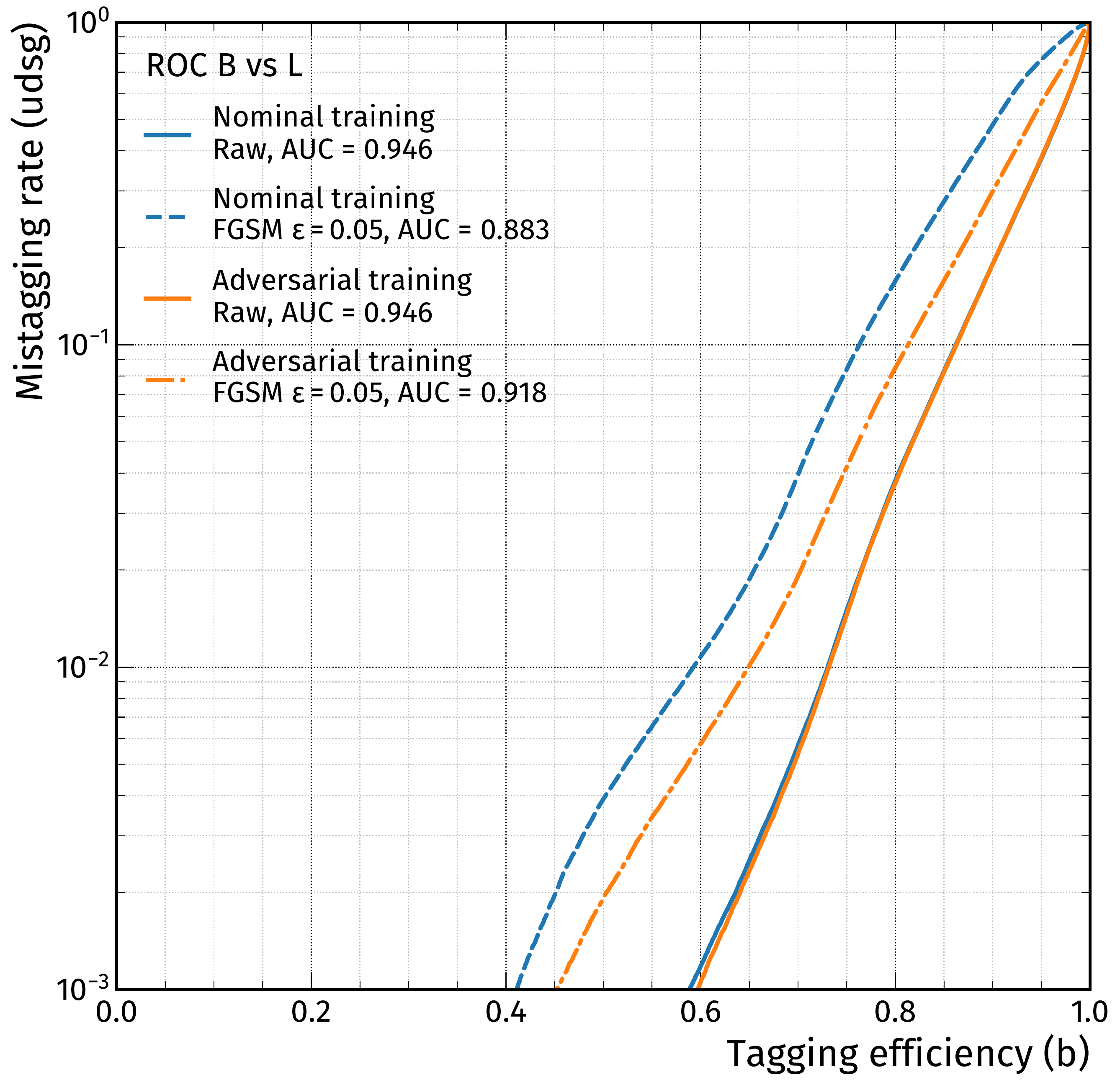} %
	\includegraphics[width=0.45\textwidth]{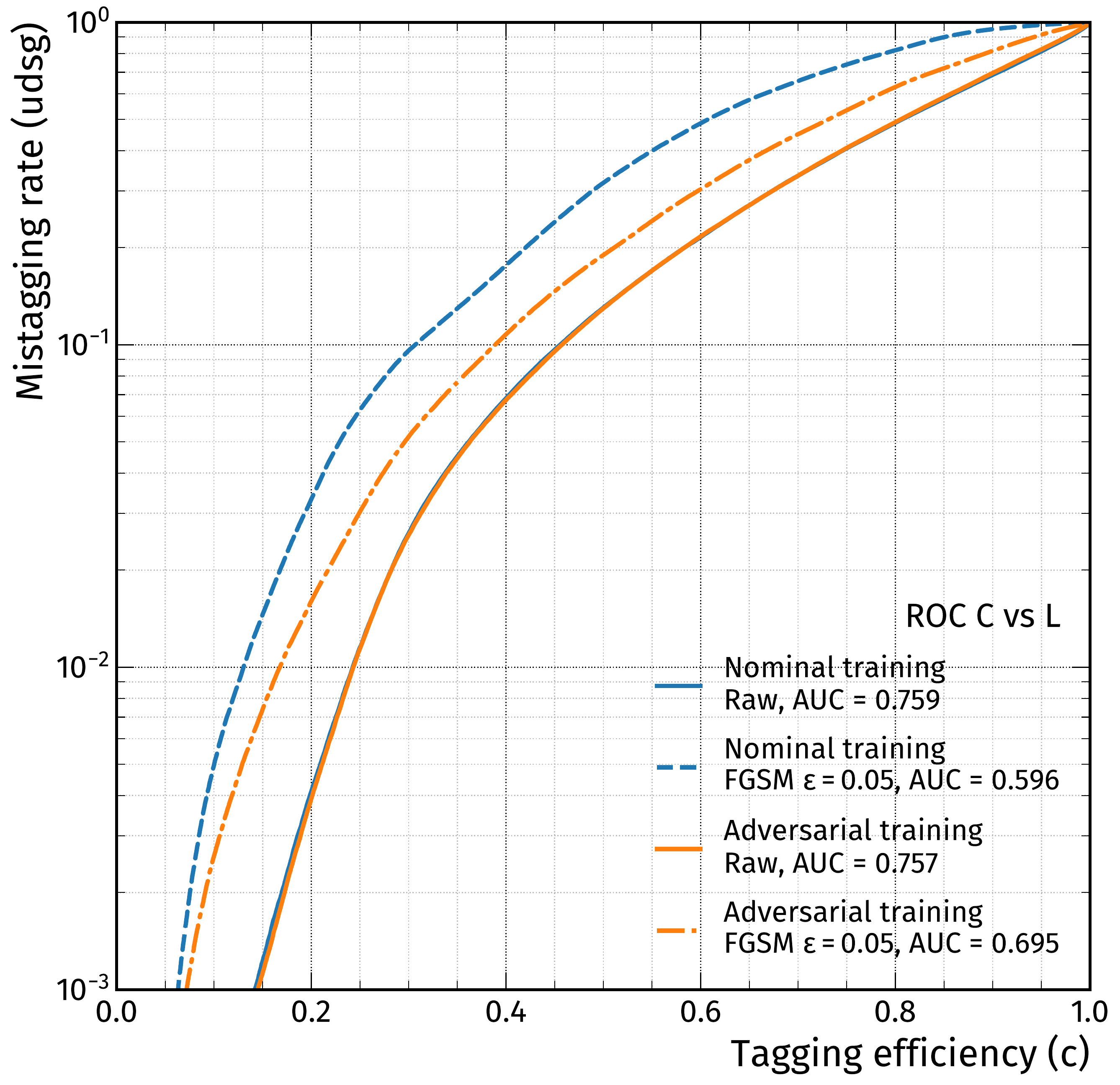}
} %
	\caption[ROC curves for the BvsL (left) and CvsL (right) discriminators, comparing the nominal with adversarial training when applying the FGSM attack to both trainings individually.]{ROC curves for the BvsL (left) and CvsL (right) discriminators, comparing the nominal with adversarial training when applying the FGSM attack to both trainings individually. Nominal training is visualized in blue, adversarial training in orange, solid lines depict nominal performance, dashed lines show performance on distorted inputs (for nominal training), dashed-dotted lines represent the systematically distorted samples for adversarial training.}%
	\label{plot:basic_adv_BvL_CvL_ROC_IND}%
\end{figure*}

As a check of robustness, we perform a direct comparison of the nominal and adversarial training, crafting the FGSM samples individually per model, with the resulting ROC curves for the BvsL and CvsL discriminators shown in Fig.~\ref{plot:basic_adv_BvL_CvL_ROC_IND}.
The corresponding AUC values for BvsL are identical ($0.946$) and are practically identical for CvsL (nominal: $0.759$, adversarial training $0.757$). At the same time, the adversarial model maintains a high performance also when given systematically distorted samples, which can be seen from the dashed-dotted lines corresponding to the colors mentioned above. The ROC curve corresponding to FGSM samples crafted for and injected to the adversarial training (orange dashed-dotted line) appears much closer to that showing nominal performance (solid line) than what can be observed for the ROC curves corresponding to the FGSM attack for the nominal training (blue lines). In numbers, this effect is best observed for the CvsL discriminator where the decrease in performance is roughly $21\%$ for the nominal training, but only $8.2\%$ for the adversarial training, while the nominal performance of both models is nearly same. Hence, we have shown that it is possible to build a more robust tagger that is simultaneously highly performant. A label leaking effect (see Ref.~\cite{kurakin2017adversarial}), which refers to a better performance on adversarial examples than on undisturbed data for an adversarial model, is not observed.

\begin{figure}[!ht]
    \centering
    \includegraphics[width=0.45\textwidth]{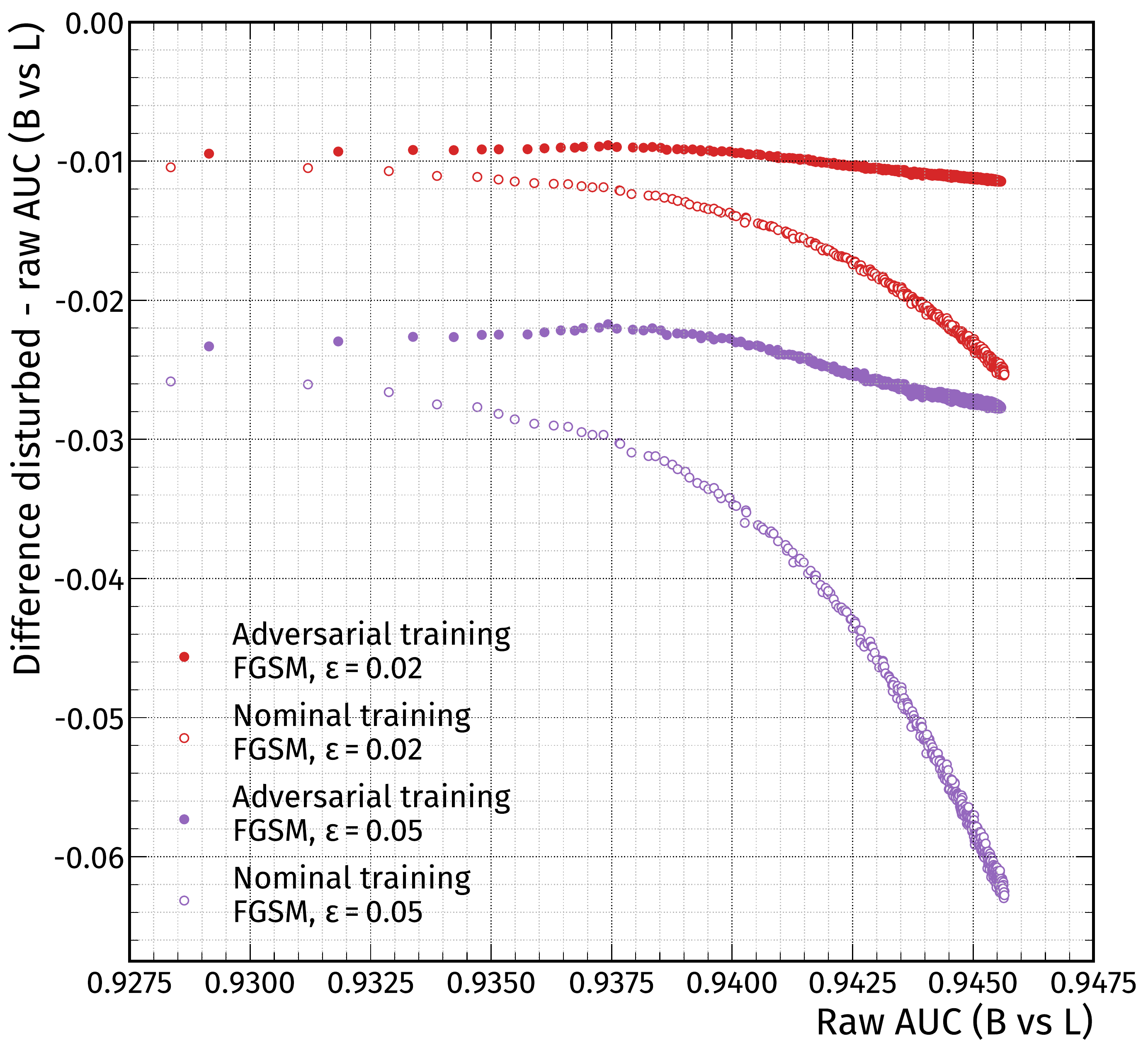}
    \caption[Relation between susceptibility and nominal performance for the nominal and adversarial training, tested on systematically distorted inputs with varying $\epsilon$ in different colors.]{Relation between susceptibility and nominal performance for the nominal and adversarial training, tested on systematically distorted inputs with varying $\epsilon$ in different colors. The $x$ axis shows nominal performance, measured with BvsL AUC, while the $y$ axis shows the difference between disturbed and raw AUC. When there is a drop on the $y$ axis while moving to higher nominal performance ($x$ axis), this indicates higher susceptibility. The empty markers represent the nominal training, which becomes highly vulnerable with increasing nominal performance (with the drop always getting steeper), while the filled markers for adversarial training show a much flatter relation.}
    \label{plot:diff_AUC_over_raw_AUC}
\end{figure}

Figure~\ref{plot:diff_AUC_over_raw_AUC} compares the susceptibility to mismodeling of the two classifiers as a function of performance. FGSM samples have been generated individually for each model and checkpoint (denoting each epoch with a single point) to scan over different discrete stages of the training. Higher density of points in the high performance region is representative of the small improvements at later stages of the training, while the performance gain during the first few epochs is quick. Ideally, there would be a constant relation that shows no signs of decreasing robustness for increasing performance. However, we observe a considerable deterioration (and thus higher susceptibility to mismodeling) of the nominal classifier. The effect for the adversarial model, while still noticeable, is to a large degree mitigated.
In fact, the adversarial training seems to recover some of its robustness (e.g. peaking at an AUC of around $0.938$) before the impact at higher performance starts to worsen the resistance. Again, this shows the intriguing trade-off between performance and robustness for the nominal training, where training to highest performance is not necessarily advisable due to high susceptibility. On the other hand, the adversarial training performs equally well on nominal samples and only shows a weak functional dependence between performance on first-order adversaries and the respective undisturbed performance.
\subsection{Probing Flavor Dependence of the Attack as a Proxy
for Generalization Capability}\label{subsec:experiments_flavor_dependence}
\begin{figure}[ht]
	\centering
	\hspace*{-0.6cm}
	\includegraphics[width=0.45\textwidth]{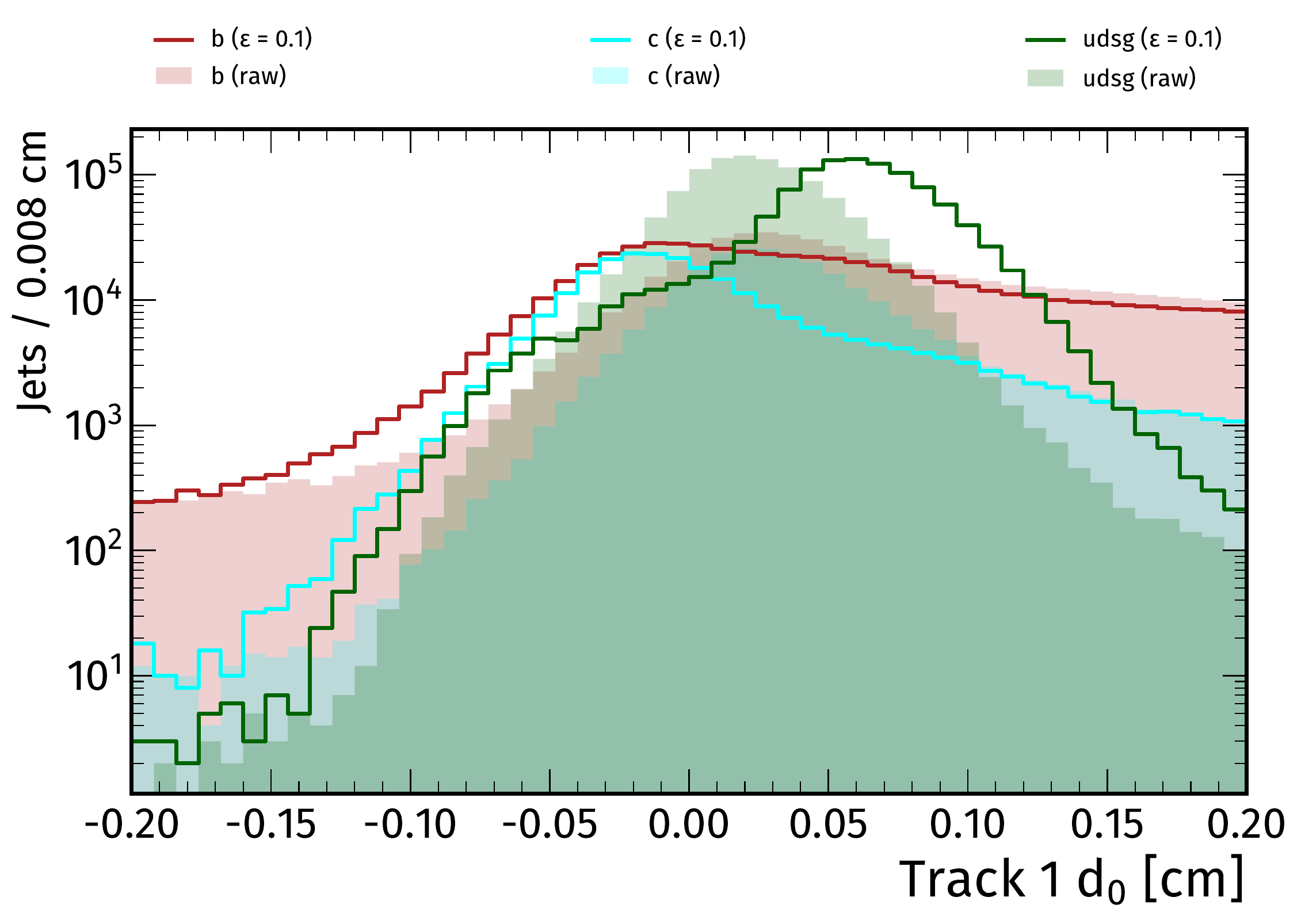}
	\hspace*{-0.6cm}
	\includegraphics[width=0.45\textwidth]{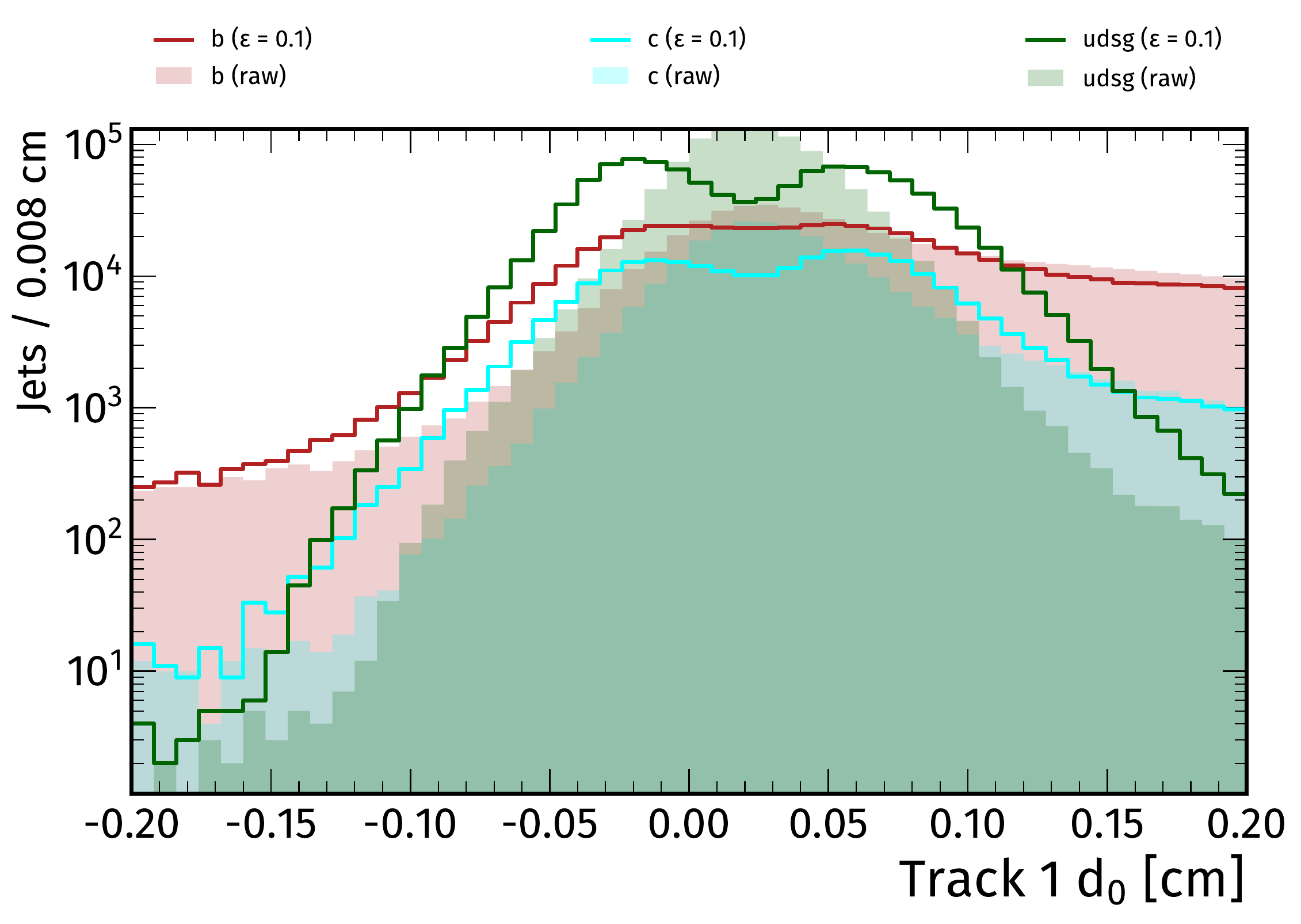}
	\caption[Signed transverse impact parameter distribution for the first track, split by flavor, before (filled histograms) and after (lines) applying the FGSM attack for the nominal (top) and adversarial (bottom) models, respectively.]{Signed transverse impact parameter distribution for the first track, split by flavor, before (filled histograms) and after (lines) applying the FGSM attack for the nominal (top) and adversarial (bottom) models, respectively. Clearly asymmetric shapes are produced when using the FGSM attack for the loss function assigned to the nominal training. Applying the FGSM attack based on an adversarial model shows suppressed flavor-dependency and relatively symmetric shapes. The attack uses the parameter $\epsilon=0.1$, which is higher than the moderately chosen parameter of $\epsilon=0.01$ during the modified training loop.}
	\label{plot:d0_splitByFlav_basic_adv_training}
\end{figure}

In an attempt to understand why the adversarial model is more robust than the nominal classifier, we investigate nominal and perturbed input distributions of a selected feature, split by flavor. We intentionally choose a large distortion. This test aims at visualizing geometric properties of the distorted samples, purposefully choosing a large $\epsilon$ of $0.1$. This is equal to the regular FGSM attack described by Eq.~(\ref{eq:fgsm}) without the limitation described in Eq.~(\ref{eq:fgsm_mod}). The signed impact parameter ($d_0$) as shown in Fig.~\ref{plot:d0_splitByFlav_basic_adv_training} originally offers discriminating power via the fact that heavy-flavor jets contain displaced tracks associated to a secondary vertex, which should naturally lead to more positive values for the $d_0$ variable. For light-flavored jets, this behaviour is not expected, instead the tracks in light jets have a roughly symmetric $d_0$ distribution, peaking at $0$, apart from some skewness due to relatively long-lived, but light hadrons ($K^0_s$ or $\Lambda$) or contamination with tracks from heavy-flavor hadrons~\cite{CMS-BTV-16-002}.

For the nominal training, light-flavor jets are shifted mostly into the positive region, which should be dominated by b jets; b jets are shifted to the negative region where these jets were not abundant previously. From a geometric point of view, the FGSM attack on the nominal training produces asymmetric shapes. On the other hand, the resulting perturbed input distributions for the adversarial training are symmetric. We observe that the adversarial model is almost agnostic to the direction into which the FGSM attack shifts the inputs, while the nominal training shows a clear preferred direction that could be described as an inversion of the expected physics. For the adversarial training, the attack seems to have difficulties deciding which direction is the worse direction, resulting in a perceived ``coin-flipping'' of the shift. Thus, the adversarial training remains less susceptible than the nominal training, even when the distortions are noticeably large.

It is conceivable that the different geometric properties of the distributions are related to the geometry of the loss surface~\cite{fawzi2016robustness,fawzi2016analysis,li2018visualizing,fort2020deep}. This is expected to be responsible for differences in robustness as well. Figure~\ref{graphic:loss_surface} illustrates how the flatness of the loss surface in the vicinity of raw inputs could influence symmetric or asymmetric shifts.

A nominal training converges into a minimum associated with the default distributions. In that case, for a given flavor, there will be a specific vector pointing away from a local minimum and the direction is fixed according to the steepest increase in loss. The adversarial training always ``sees'' (new) adversarial inputs, so the adjustment of the model's parameters might average out eventually over further training epochs. Always following the newly distorted inputs yields a locally constant loss manifold around the original inputs due to the more complex saddle point problem. This would mean that not the exact memorization of training data, but rather higher-order correlations contribute to the improvement of the performance of the adversarial training~\cite{chakraborty2018adversarial,madry2019deep,fawzi2016robustness,fawzi2016analysis}. With the assumption of a flat loss surface close to the raw inputs there would be no preferred direction for first-order adversarial attacks crafted for the adversarial model. Many vectors would fulfill the criterion of pointing in the direction of increasing loss, much like choosing the direction randomly. 

Thus by examining the geometric properties of adversarial samples, a flat loss landscape for the adversarial model is highly probable, leading to higher robustness~\cite{fawzi2016robustness,fawzi2016analysis}. For mismodelings of order $\epsilon$ that are still on-manifold, the adversarial training would generalize better to data than nominal training. Robustness and generalization are not equivalent~\cite{chakraborty2018adversarial,madry2019deep,stutz2019disentangling}, which is why the above statement can not be general, but is only valid under the assumption that adversarial methods like the FGSM attack replicate mismodelings between simulation and detector data.
\section{Conclusion}\label{sec:conclusion}
In this paper, we investigated the performance of a jet flavor tagging algorithm when being exposed to systematically distorted inputs that have been generated with an adversarial attack, the Fast Gradient Sign Method. Moreover, we showed how model performance and robustness are related. We explored the trade-off between performance on unperturbed and on distorted test samples, investigating ROC curves and AUC scores for the BvsL and CvsL discriminators. All tests conducted with the nominal training confirm earlier findings that relate higher performance with higher susceptibility, now for a deep neural network that replicates a typical jet tagging algorithm. We applied a defense strategy to counter first-order adversarial attacks by injecting adversarial samples already during the training stage of the classifier, but without altering the network architecture.

When comparing this new classifier with the nominal model, no difference in performance was observed, but the robustness towards adversarial attacks is enhanced by a large margin. Exemplary for the direct comparison of the two trainings, both reached an AUC score of approximately $76\%$ when discriminating c from light jets, but an FGSM attack that is still moderate in its impact on the input distributions decreases the performance of the nominal training by $21\%$, and only by $8.2\%$ for the adversarial training. A study of raw and distorted input distributions allowed us to relate geometric properties of the attack with geometric properties of the underlying loss surfaces for a nominal and an adversarially trained model, yielding a possible explanation for the higher robustness of the latter attributed to flatness of the loss manifold.

To some extent, the higher robustness as shown in this paper points at better generalization capability, but a study that will also utilize detector data has yet to be conducted to confirm this conjecture. The approach followed for this work is comparatively general, in that it only needs access to the model and the criterion. This is the first application of adversarial training to build a robust jet flavor tagger suitable for usage at the LHC.

It would be interesting to apply this type of attack and defense also to more complex neural network structures to see if, for example, convolutional layers are able to leverage adversarial attacks differently, and if adversarial training is as effective for taggers with a larger (or smaller) dimension in the feature space. Another focus could be targeted at using adversarial methods of higher complexity, both for the attack, as well as for the defense against them. Summarizing the efforts so far, adversarial training was applied successfully to resist first-order adversarial attacks on jet flavor tagging algorithms, corresponding studies with higher-order adversaries are left for future investigations.

\begin{acknowledgements}
Simulations were performed with computing resources granted by RWTH Aachen University under project \texttt{nova0021} and \texttt{rwth0619}. This work has received support by the Deutsche Forschungsgemeinschaft (DFG, German Research Foundation, projects SCHM 2796/5 and GRK 2497), and the Bundesministerium für Bildung und Forschung (BMBF, Project 05H2021). We thank Nicolas Frediani for his contributions to the project in context of his bachelor thesis. 
\end{acknowledgements}

\section*{Declarations}
\paragraph{Data Availability Statement}
{\footnotesize This manuscript has associated data in a data repository [Authors’ comment: The dataset has been generated with code accessible under Ref.~\cite{delphes-rave} and can be accessed at the UCI Machine Learning in Physics Web portal under the link \mbox{\url{http://mlphysics.ics.uci.edu/}}].}

\paragraph{Code Availability}
{\footnotesize Results shown in this report have been prepared with the help of code accessible under Ref.~\cite{annikastein_code_github}.}

\paragraph{Funding}
{\footnotesize Open Access funding enabled and organized by Projekt DEAL.}

\paragraph{Conflict of Interest}
{\footnotesize On behalf of all authors, the corresponding author states that there is no conflict of interest.}

\bibliographystyle{my_unsrt}
\bibliography{citations_unsrt}   

\appendix
\section{Supplementary Material}\label{sec:app-supp}
\begin{figure}[!ht]%
	\centering
	\includegraphics[width=0.45\textwidth]{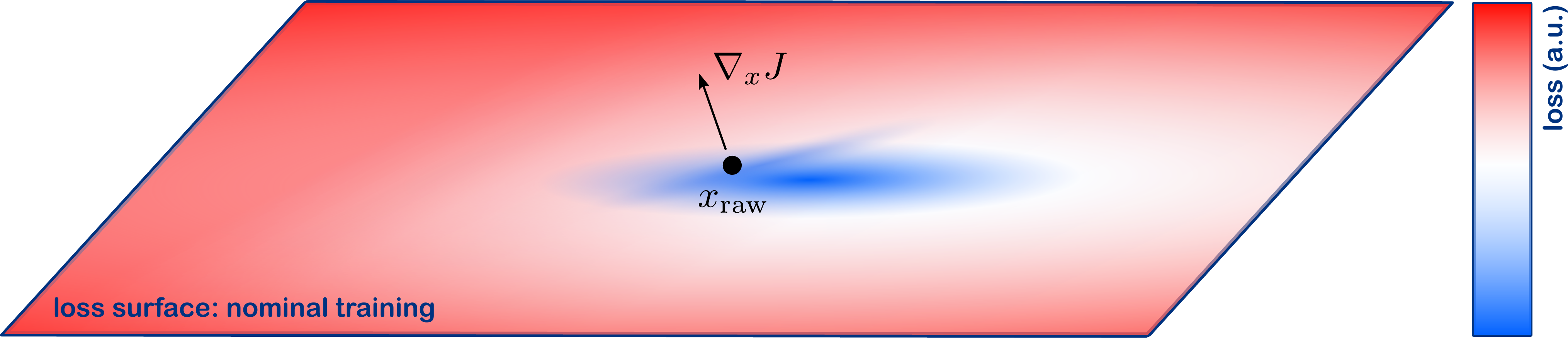}\\ %
	\vspace{0.3cm}
	\includegraphics[width=0.45\textwidth]{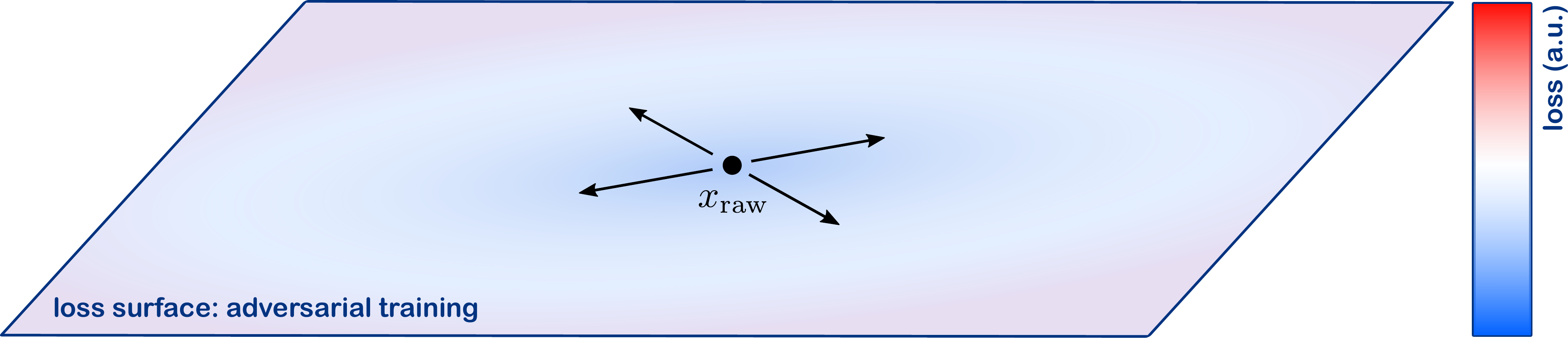} %
	\caption[Illustration of the potential geometry of the loss surfaces for the nominal as well as the adversarial training.]{Illustration of the potential geometry of the loss surfaces for the nominal as well as the adversarial training. Inspired by Refs.~\cite{fawzi2016robustness,fawzi2016analysis,li2018visualizing}.}%
	\label{graphic:loss_surface}%
\end{figure}

See Fig.~\ref{graphic:loss_surface}.
\section{Robustness in the Context of Other
Mismodeling Scenarios}\label{sec:mismodeling_scenarios}
\subsection{Smearing Inputs with a Gaussian Noise Term}\label{subsec:gaussian_smearing}
While the FGSM attack aims at worst-case scenarios in the direction of increasing gradients, a physical effect induced by mismodeling of the parton shower or caused by detector misalignment or -calibration does not know the model parameters or its loss surface. It can therefore not act as a ``demon''~\cite{nachman2019ai} that always points in a preferred direction. Investigating a smearing technique independent of the model under consideration is of interest when studying the robustness to more typical mismodeling scenarios or fluctuations that are of statistical nature. A non-systematic strategy to create a new, slightly distorted set of inputs randomly shifts the variables by adding a noise term $\xi$ to the original inputs, drawn from a Gaussian distribution~\cite{Goodfellow-et-al-2016,szegedy2014intriguing,goodfellow2015explaining,kurakin2017adversarial}:
\begin{align}
	x_\text{noise} = x_\text{raw} + \xi, \qquad \text{where}
\end{align}
\begin{align}
	\xi \sim \mathcal{N}\left(\mu,\sigma^2\right) \qquad
	\text{with}\qquad P(\xi)=\frac{1}{\sigma \sqrt{2\pi}} e^{-\frac{1}{2}\left(\frac{\xi-\mu}{\sigma}\right)^2}.
\end{align}
As described in Sect.~\ref{subsec:preprocessing}, the inputs are scaled to a standard deviation of one and are centered at zero, thus allowing this smearing without further processing.

\begin{figure}[ht]%
	\centering
	\includegraphics[width=0.45\textwidth]{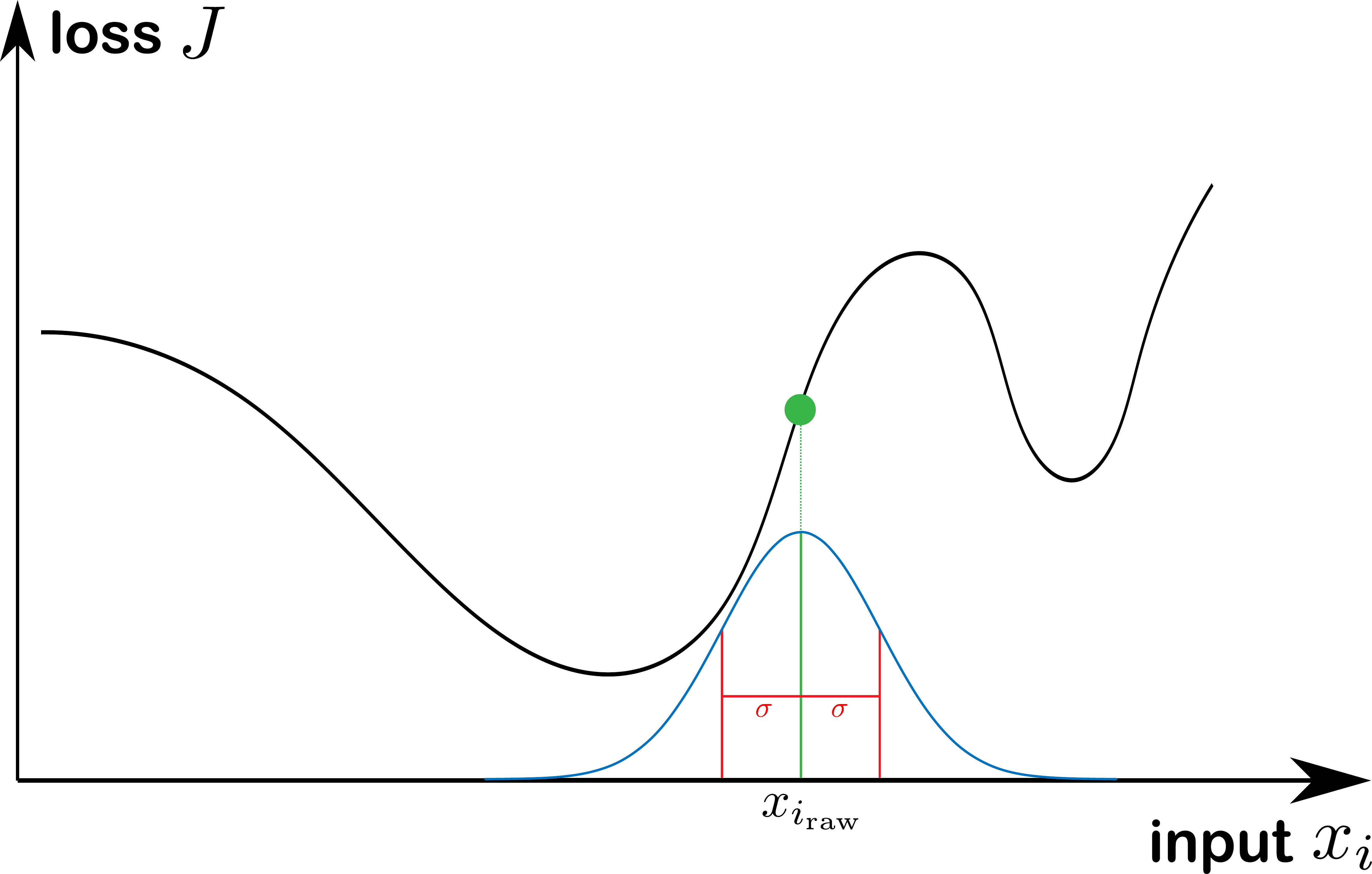} %
	\caption[Visualization of the random shift of inputs by adding a Gaussian noise term.]{Visualization of the random shift of inputs by adding a Gaussian noise term. With the slight distortion based on the blue probability distribution, the formerly green raw datapoint is shifted and the corresponding loss modified. The change of the loss function with respect to the distorted inputs can go in either direction. Gaussian distribution adapted from Ref.~\cite{gaussian}.}%
	\label{graphic:noise}%
\end{figure}

The effect of this distortion is shown in Fig.~\ref{graphic:noise}. Only one arbitrary input $x_i$ has been chosen for visualization, and the displayed loss function is just an illustration.
Compared to the settings introduced for the FGSM attack, the difference is that the magnitude of the distortion is now given by $\sigma=1$ (not $\epsilon$). Other parameters remain untouched, the limitation by $25\%$ of the value applies as well and we choose $\mu=0$ for this test against random fluctuations of features.
\begin{figure*}[ht]%
	\makebox[\textwidth][c]{
	\centering
	\includegraphics[width=0.45\textwidth]{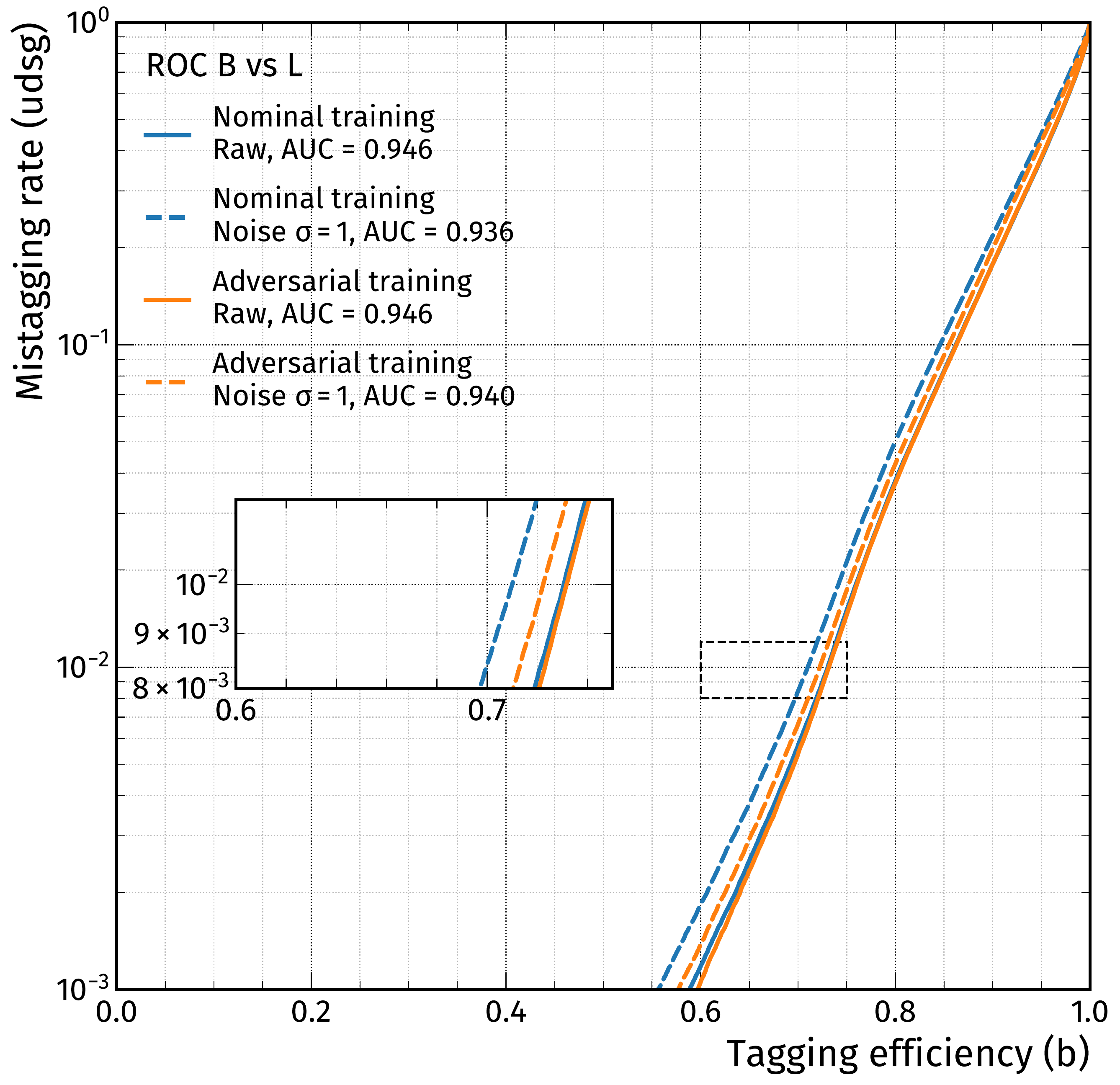} %
	\includegraphics[width=0.45\textwidth]{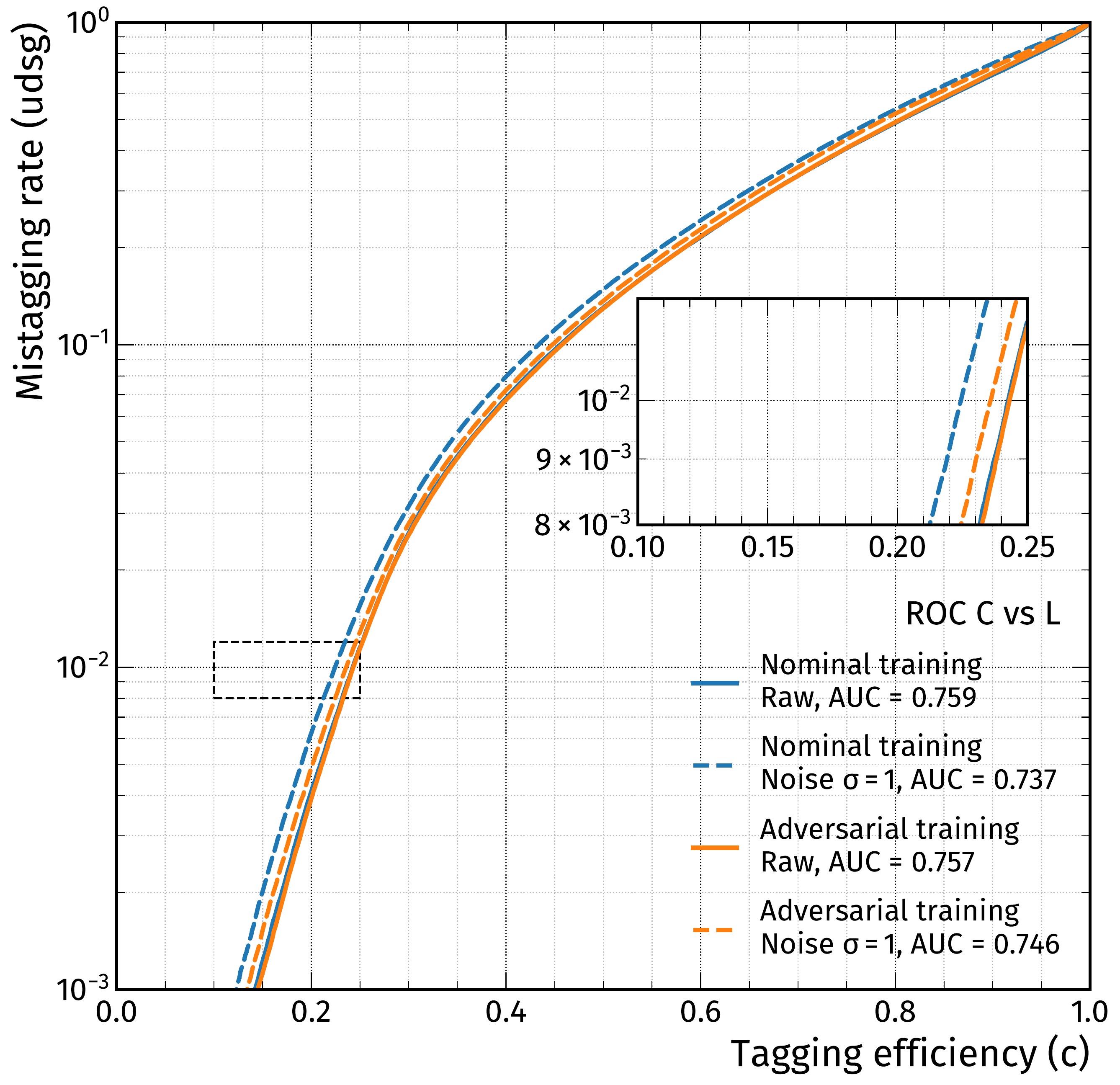}
} %
	\caption[ROC curves for the BvsL (left) and CvsL (right) discriminators, comparing the nominal with adversarial training when smearing the inputs with a Gaussian noise term.]{ROC curves for the BvsL (left) and CvsL (right) discriminators, comparing the nominal with adversarial training when smearing the inputs with a Gaussian noise term. Nominal training is visualized in blue, adversarial training in orange, solid lines depict nominal performance, dashed lines show performance on distorted inputs.}%
	\label{plot:basic_adv_BvL_CvL_ROC_NOISE}%
\end{figure*}
From Fig.~\ref{plot:basic_adv_BvL_CvL_ROC_NOISE} it is evident that the adversarial model also performs better than the nominal model when tested on randomly smeared inputs, although the advantage over nominal training is not as large as for the FGSM attack. Measured with difference in AUC, adversarial training brings a factor of 2 smaller susceptibility to Gaussian noise, compared to nominal training. Therefore we conclude that also in this scenario, which is somewhat closer to typical mismodelings found in the HEP context, the adversarial training is more robust.
\subsection{Transferability of Adversarial Samples
as a Black‑Box Attack}\label{subsec:transferability}
\begin{figure*}[ht]%
	\makebox[\textwidth][c]{
	\centering
	\includegraphics[width=0.45\textwidth]{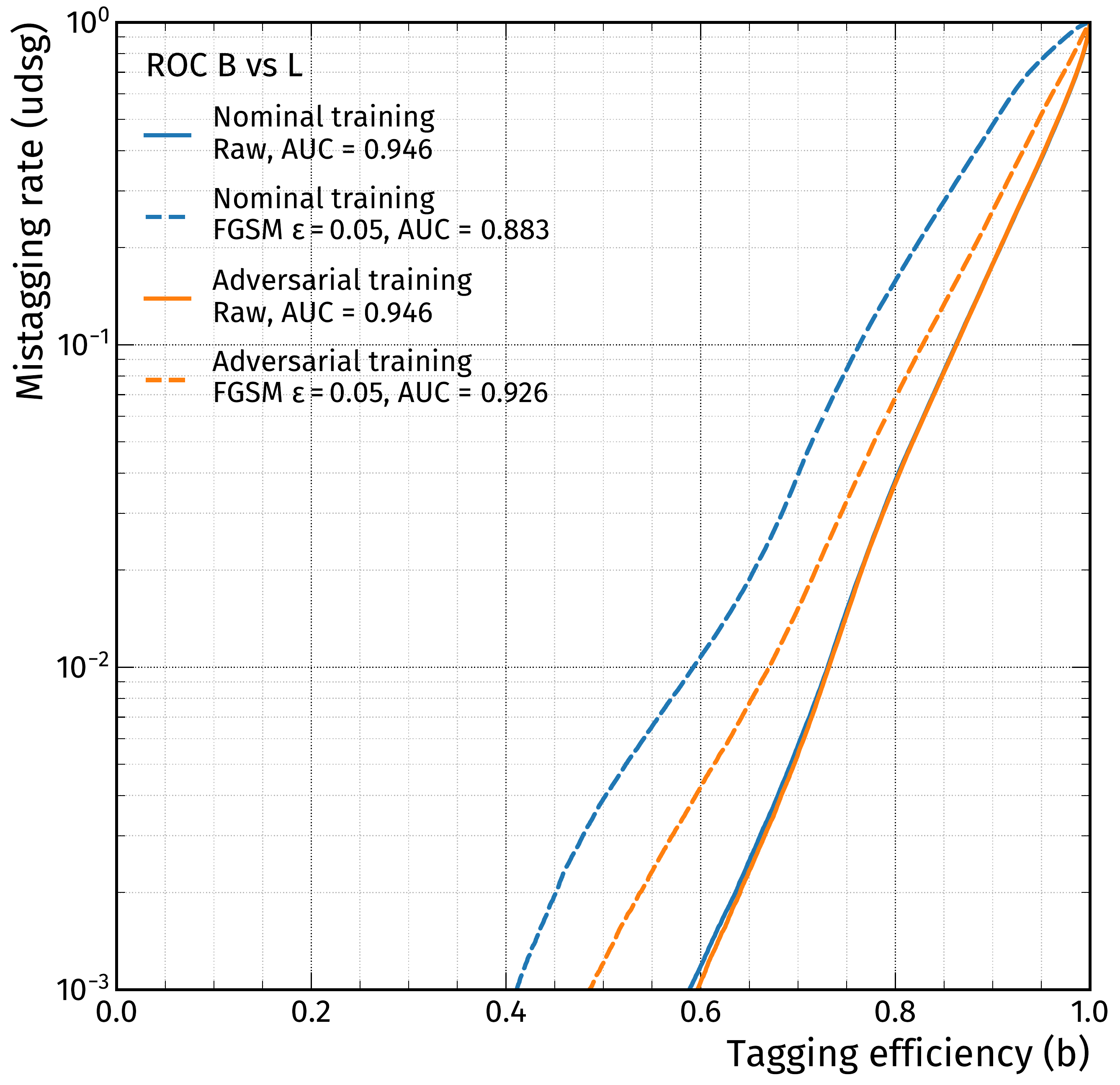} %
	\includegraphics[width=0.45\textwidth]{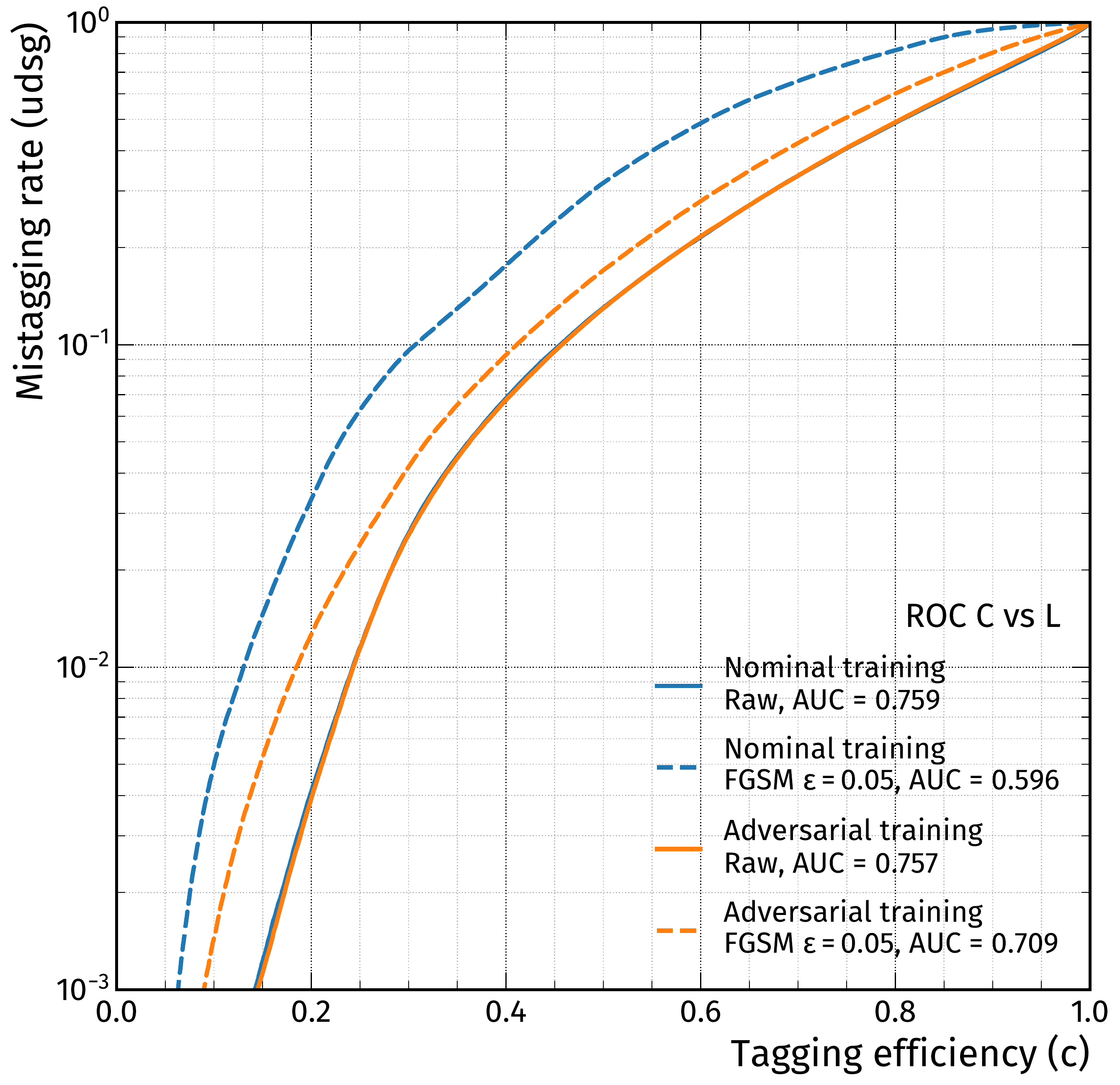}
} %
	\caption[ROC curves for the BvsL (left) and CvsL (right) discriminators, comparing the nominal with adversarial training when applying the FGSM attack to the nominal training and injecting the obtained inputs to both models.]{ROC curves for the BvsL (left) and CvsL (right) discriminators, comparing the nominal with adversarial training when applying the FGSM attack to the nominal training and injecting the obtained inputs to both models. Nominal training is visualized in blue, adversarial training in orange, solid lines depict nominal performance, dashed lines show performance on distorted inputs that were obtained with the help of the loss surface of the nominal model.}%
	\label{plot:basic_adv_BvL_CvL_ROC_FROMBASIC}%
\end{figure*}

Adversarial samples created for one model can also deteriorate the performance of another, independent model, which is known as transferability of adversarial samples~\cite{goodfellow2015explaining,chakraborty2018adversarial,madry2019deep}. For this study, the two models under consideration share the same architecture, but the weights and bias terms differ as a result of the different training strategies, thus yielding suiting candidates to investigate the aforementioned transferability. In fact, when injecting the same FGSM inputs generated for the nominal model into both models, we obtain another set of predictions. This can be understood as a black-box attack on the adversarial model~\cite{madry2019deep}, as the adversarial inputs are crafted without knowledge of the exact parameters of the adversarial model. In Fig.~\ref{graphic:inference_nominal_adversarial}, this corresponds to using the blue branch for both models as an identical set of samples. The parameter used for this scenario is $\epsilon=0.05$, with the limitation introduced in Eq.~(\ref{eq:fgsm_mod}). Figure~\ref{plot:basic_adv_BvL_CvL_ROC_FROMBASIC} shows that the adversarial model is also more robust to this perturbation.
\subsection{Shifting Inputs Systematically with Up/Down
Variations}\label{subsec:systematic_up_down}
In this simplified scenario, inputs are modified without prior knowledge of the model parameters. For this variation, features are simultaneously shifted upwards (or downwards) by adding (subtracting) small distortions to (from) the nominal values. Whereas the present dataset does not contain the systematic uncertainties directly, we estimate the magnitude of the distortion that is applied in a feature-wise manner with the help of existing commissioning results~\cite{CMS-BTV-16-002,ATLASbtagging,CMS-PAS-BTV-16-001,ATLASctagging} by the CMS and ATLAS collaborations. A baseline magnitude of $0.05$ has been chosen, which is weighted by a factor $s_i$ ranging from $1$ to $5$, depending on the maximally observed data-to-simulation disagreement for input $i$:
\begin{align}\label{eq:syst}
	x_{\text{sys},i} = x_{\text{raw},i} \pm 0.05\cdot s_i, \quad \text{where} \quad s_i\in \{1,\dots,5\}.
\end{align}
\begin{figure*}[ht]%
	\makebox[\textwidth][c]{
	\centering
	\includegraphics[width=0.45\textwidth]{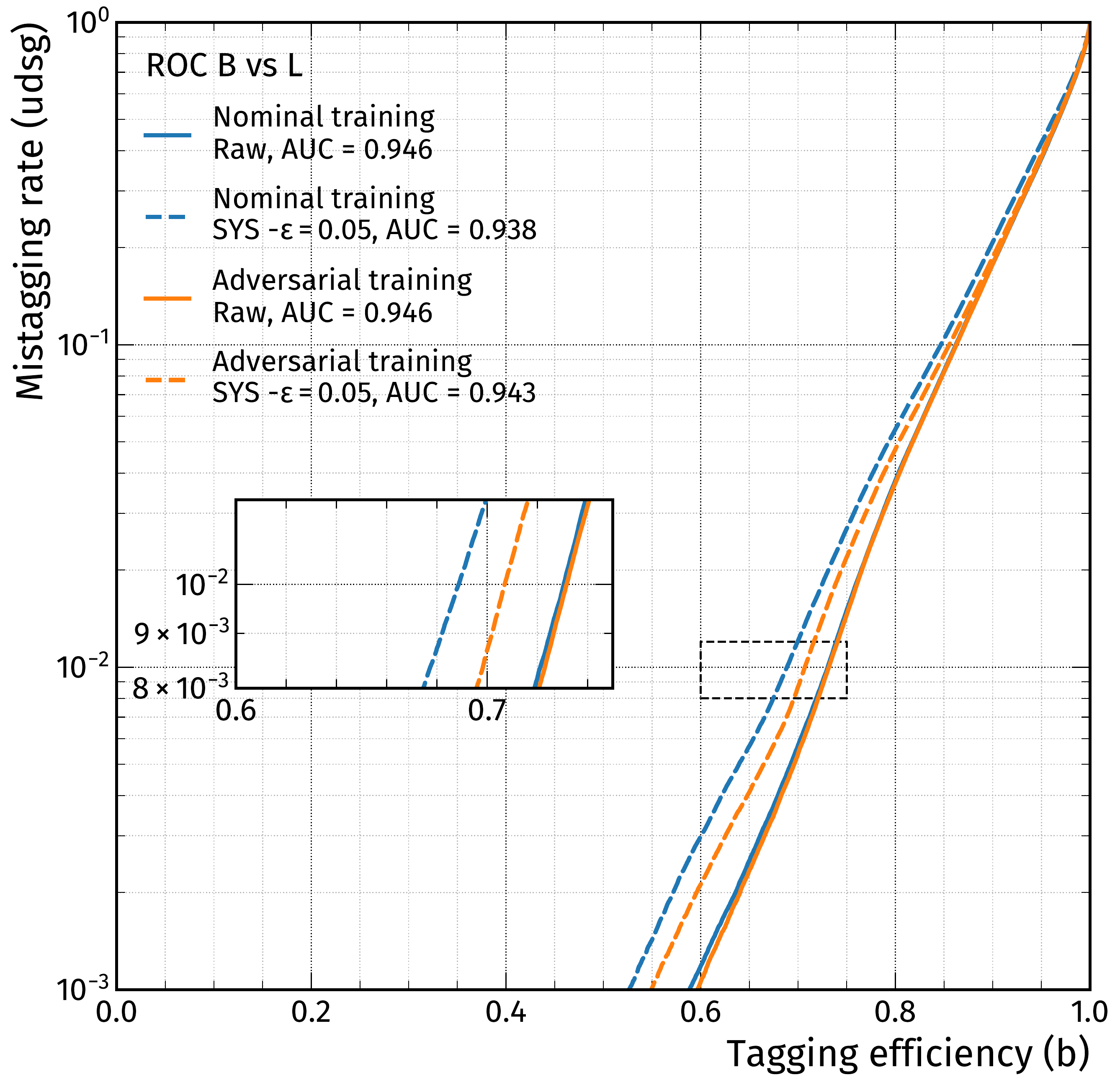} %
	\includegraphics[width=0.45\textwidth]{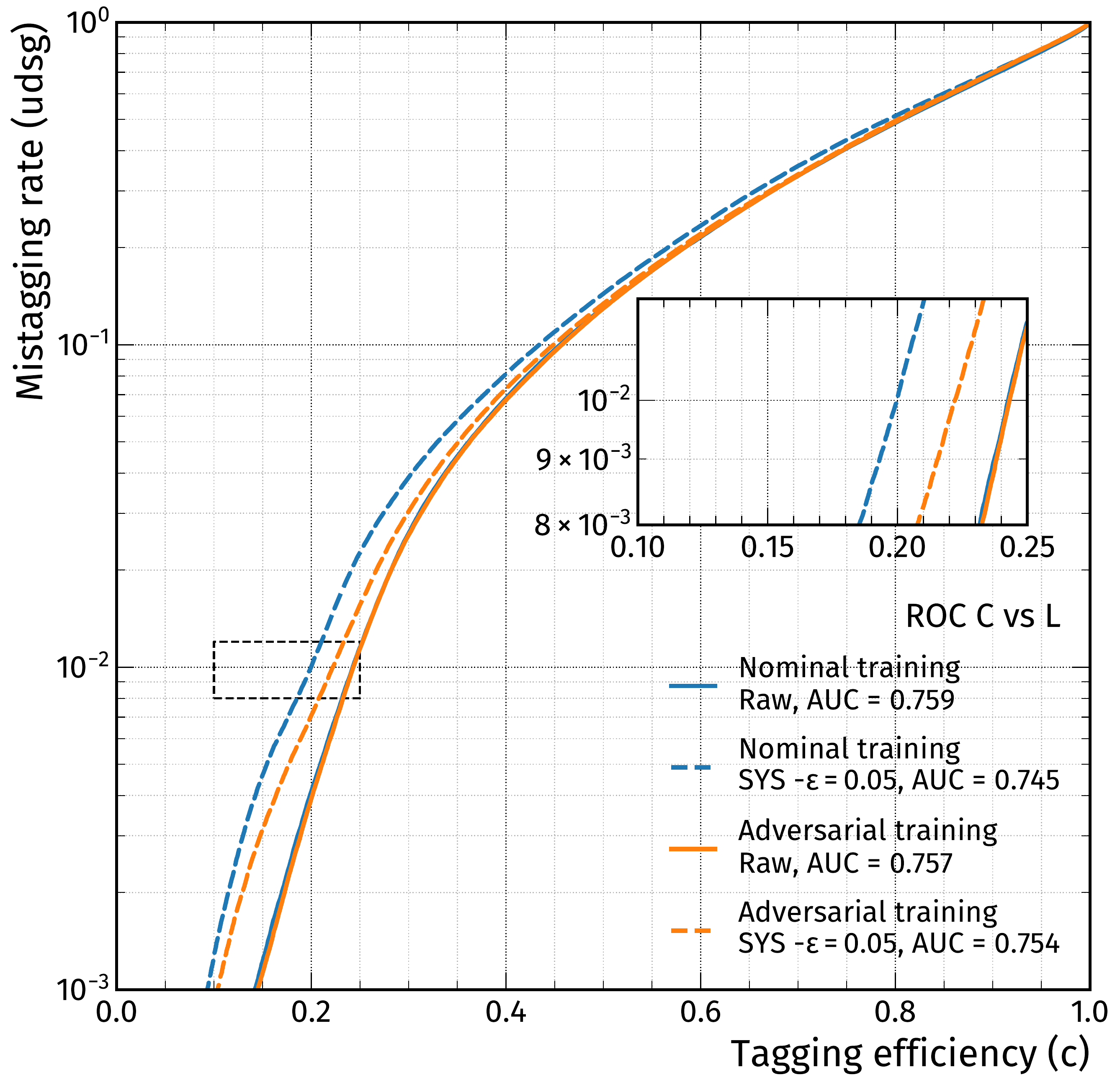}
} %
	\caption[ROC curves for the BvsL (left) and CvsL (right) discriminators, comparing the nominal with adversarial training when shifting inputs systematically downwards.]{ROC curves for the BvsL (left) and CvsL (right) discriminators, comparing the nominal with adversarial training when shifting inputs systematically downwards. Nominal training is visualized in blue, adversarial training in orange, solid lines depict nominal performance, dashed lines show performance on distorted inputs.}%
	\label{plot:basic_adv_BvL_CvL_ROC_DOWN}%
\end{figure*}
\begin{figure*}[ht]%
	\makebox[\textwidth][c]{
	\centering
	\includegraphics[width=0.45\textwidth]{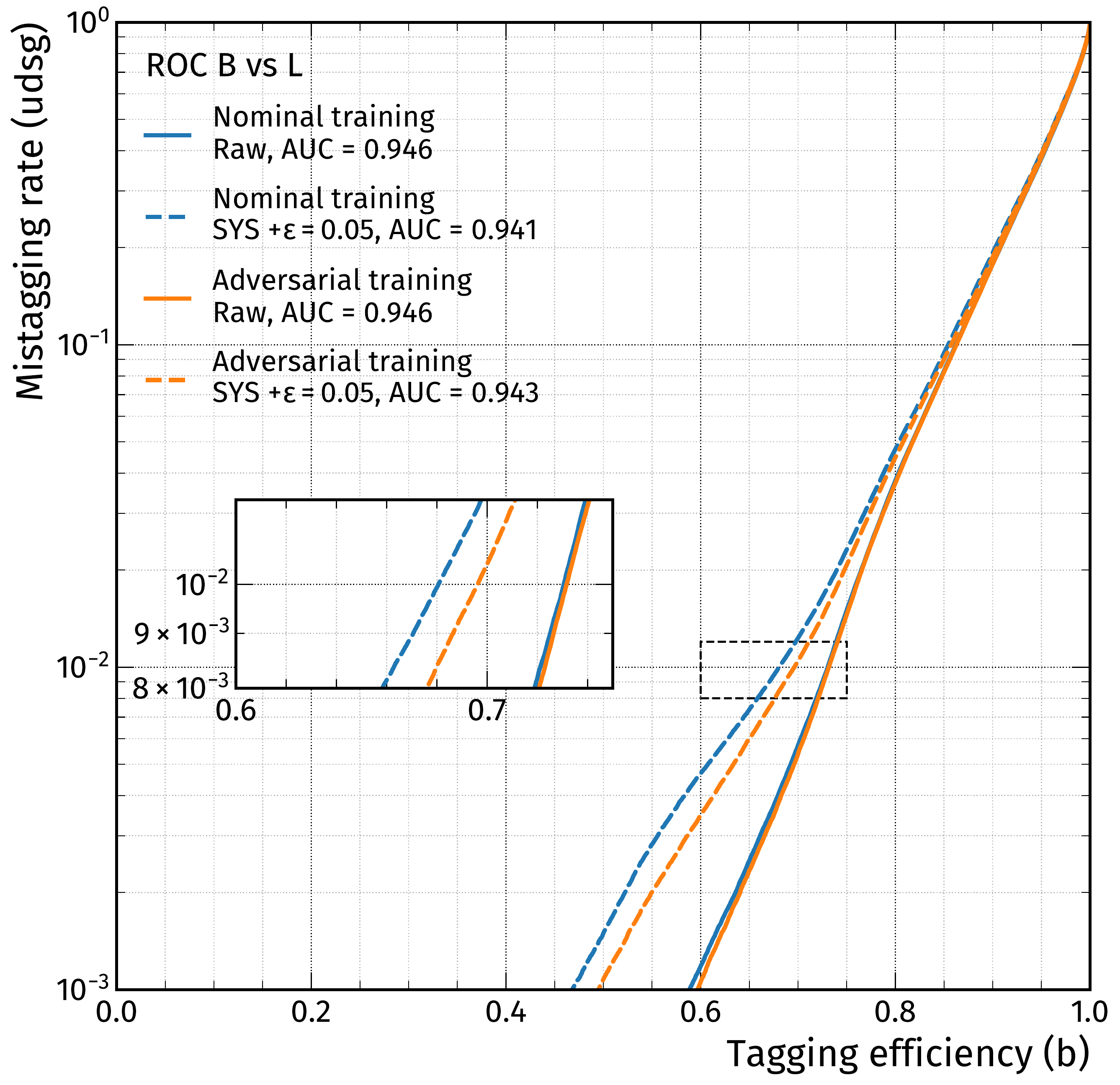} %
	\includegraphics[width=0.45\textwidth]{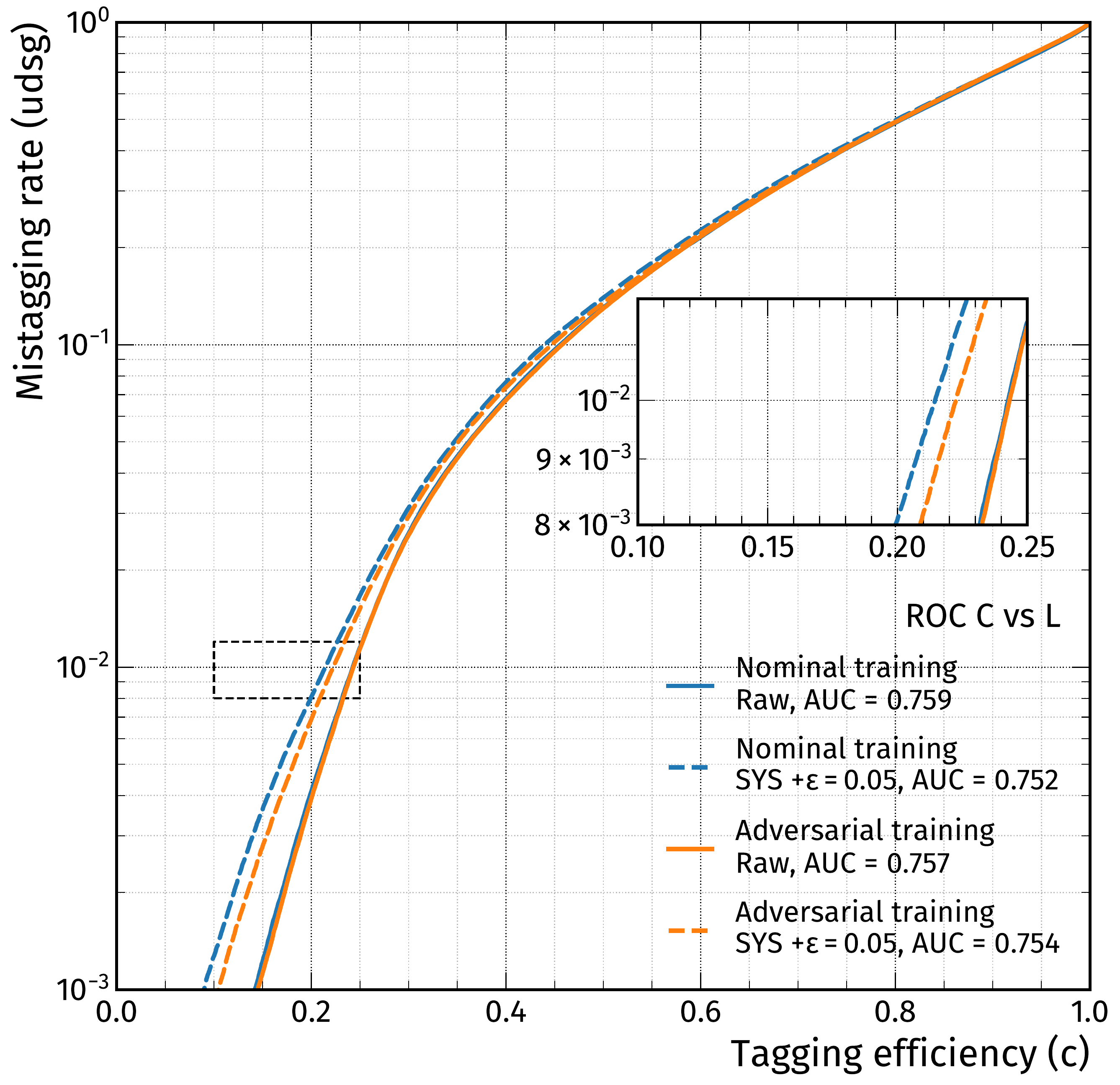}
} %
	\caption[ROC curves for the BvsL (left) and CvsL (right) discriminators, comparing the nominal with adversarial training when shifting inputs systematically upwards.]{ROC curves for the BvsL (left) and CvsL (right) discriminators, comparing the nominal with adversarial training when shifting inputs systematically upwards. Nominal training is visualized in blue, adversarial training in orange, solid lines depict nominal performance, dashed lines show performance on distorted inputs.}%
	\label{plot:basic_adv_BvL_CvL_ROC_UP}%
\end{figure*}
The largest deviation in the data-to-simulation ratio is accounted for by incrementing the initial factor of $1$ $s_i$-times in steps of $1$, where $s_i$ counts how many intervals of $0.1$ fit between the observed ratio and perfect agreement (i.e. a ratio of $1$). This already introduces a restriction on the allowed perturbation by itself, which is why the additional limitation of $25\%$ is not necessary here. For features in the dataset where no direct counterpart is used in the official taggers of said collaborations, or data-to-simulation comparisons are unavailable, reasonable intermediate factors are assumed in Eq.~(\ref{eq:syst}).

Figures~\ref{plot:basic_adv_BvL_CvL_ROC_DOWN} and~\ref{plot:basic_adv_BvL_CvL_ROC_UP} prove that in the case of simultaneous up- or downwards variations, the adversarial model maintains a higher performance than the nominal model. The impact of this distortion is not as large as the one observed for the FGSM attack and further, this perturbation does not take correlations into account, which is why the advantage of adversarial training over nominal training is not as enhanced and we might not have seen the worst possible case yet. However, in this simplified scenario, adversarial training can be considered as more robust towards systematical shifts of input features. 
\section{Computing}\label{sec:app-computing}
Processing of the data is carried out with the \verb|awkward|~\cite{jim_pivarski_2020_4341376} package, later evaluation is facilitated by utilizing \verb|coffea|~\cite{lindsey_gray_2020_3266454}, the graphics are prepared with \verb|matplotlib|~\cite{mpl}. The neural network training is performed with the \verb|PyTorch|~\cite{paszke2019pytorch} library, where a NVIDIA Tesla V100 GPU is utilized.
\section{Input Variables}\label{sec:app-input_variables}
See Tables~\ref{tab:highlevel_inputs},\ref{tab:lowlevel_inputs}.
\begin{table*}[ht]
\begin{center}
\begin{minipage}{\textwidth}
\caption{Expert / high-level features for the neural network. The first two features represent high-level jet, the next four high-level track, and the remaining ones high-level vertex variables. Adapted from Refs.~\cite{Guest_2016,delphes-rave}.}\label{tab:highlevel_inputs}%
\begin{tabular}{p{0.25\textwidth}p{0.65\textwidth}}
\hline\noalign{\smallskip}
Short name & Description \\
\noalign{\smallskip}\hline\noalign{\smallskip}
		{\footnotesize Jet $p_\mathrm{T}$} & {\footnotesize Transverse momentum of the jet with respect to the beam line}\\
		{\footnotesize Jet $\eta$} & {\footnotesize jet pseudorapidity}\\\noalign{\smallskip}\hline\noalign{\smallskip}
		{\footnotesize Track 2 (3) $d_0$ ($z_0$) significance} & {\footnotesize Magnitude of impact parameter significance of the second (third) track, transverse to the (along the) beam line, after ranking them by $|d_0|$ significance}\\
		{\footnotesize N tracks over $d_0$ threshold} & {\footnotesize Number of tracks with transverse impact parameter significance over $1.8$}\\
		{\footnotesize Jet Prob} & {\footnotesize Light jet probability (see Ref.~\cite{jetprob}); product of likelihoods over all tracks to have come from a light quark jet}\\
		{\footnotesize Jet width $\eta$ ($\phi$)} & {\footnotesize Width of the jet in $\eta$ ($\phi$) coordinates, obtained from all tracks in the jet via
		\begin{gather*}
		    \left(\frac{\sum_i p_{\text{T},i} \Delta\eta_i^2}{\sum_i p_{\text{T},i}}\right)^{1/2} \quad \left(\text{or} \left(\frac{\sum_i p_{\text{T},i} \Delta\phi_i^2}{\sum_i p_{\text{T},i}}\right)^{1/2}\right) \quad\text{with respect to the jet axis}
		\end{gather*}
		}\\\noalign{\smallskip}\hline\noalign{\smallskip}
		{\footnotesize Vertex significance} & {\footnotesize Weighted sum over displacement significances for all secondary vertices in the jet 
		\begin{gather*}
		    \frac{\sum_i d_i/\sigma_i^2}{\sqrt{\sum_i 1/\sigma_i^2}}
		\end{gather*}
		}\\
		{\footnotesize N secondary vertices} & {\footnotesize Number of reconstructed secondary vertices in the jet}\\
		{\footnotesize N secondary vertex tracks} & {\footnotesize Number of tracks associated to a secondary vertex, summed over all secondary vertices in the jet}\\
		{\footnotesize Vertex $\Delta R$} & {\footnotesize Sum of angular separation between the secondary vertices and the jet, weighted by the number of tracks at the SV
		\begin{gather*}
		    \frac{1}{\sum_i \text{SV}_i\text{\_nTracks}} \sum_i \text{SV}_i\text{\_nTracks} \cdot \Delta R (\text{Jet}, \text{SV}_i)
		\end{gather*}
		}\\
		{\footnotesize Vertex mass} & {\footnotesize Decay chain mass, i.e. sum over all secondary vertex masses in the jet, under the pion mass hypothesis for reconstructed particles}\\
		{\footnotesize Vertex energy fraction} & {\footnotesize Summed fractions of total track energy in the jet associated to secondary vertices}\\
\noalign{\smallskip}\hline
\end{tabular}
\end{minipage}
\end{center}
\end{table*}
\begin{table*}[ht]
\begin{center}
\begin{minipage}{\textwidth}
\caption{Low-level and intermediate-level features for the neural network. The first seven features represent low-level tracking, the remaining ones intermediate-level vertexing variables. These quantities are used for up to six tracks, ranked by signed transverse impact parameter significance. Vertex information is stored on a per-track level by utilizing a $N\rightarrow 1$ mapping. Adapted from Refs.~\cite{Guest_2016,delphes-rave}.}\label{tab:lowlevel_inputs}%
\begin{tabular}{p{0.3\textwidth}p{0.6\textwidth}}
\hline\noalign{\smallskip}
Short name & Description \\
\noalign{\smallskip}\hline\noalign{\smallskip}
		{\footnotesize Track $d_0$} & {\footnotesize Impact parameter of the track, transverse to the beam line}\\
		{\footnotesize Track $z_0$} & {\footnotesize Impact parameter of the track, along the beam line}\\
		{\footnotesize Track $\phi$} & {\footnotesize Azimuthal angle with respect to the beam axis}\\
		{\footnotesize Track $\theta$} & {\footnotesize Polar angle with respect to the beam axis}\\
		{\footnotesize Track $Q/p$} & {\footnotesize Charge over momentum}\\
		{\footnotesize Covariance between helix parameters} & {\footnotesize All $15$ independent entries of the symmetric $5\times5$ covariance matrix between helix parameters of the track}\\
		{\footnotesize Track weight} & {\footnotesize Probability for a track to be associated with the primary vertex}\\\noalign{\smallskip}\hline\noalign{\smallskip}
		{\footnotesize Vertex mass} & {\footnotesize Invariant mass of constituents used in secondary vertex fit}\\
		{\footnotesize Vertex displacement} & {\footnotesize Displacement of secondary vertex in transverse direction with respect to the interaction point}\\
		{\footnotesize Vertex displacement significance} & {\footnotesize Secondary vertex displacement divided by uncertainty of that displacement}\\
		{\footnotesize N tracks} & {\footnotesize Number of tracks associated to the secondary vertex}\\
		{\footnotesize Vertex-Jet $\Delta \eta$} & {\footnotesize Angular separation in $\eta$ between jet axis and secondary vertex}\\
		{\footnotesize Vertex-Jet $\Delta \phi$} & {\footnotesize Angular separation in $\phi$ between jet axis and secondary vertex}\\
		{\footnotesize Vertex energy fraction} & {\footnotesize Fraction of jet energy carried by tracks associated to the secondary vertex}\\
\noalign{\smallskip}\hline
\end{tabular}
\end{minipage}
\end{center}
\end{table*}

\end{document}